# An "Improved" Lattice Study of Semi-leptonic Decays of D-Mesons

*UKQCD Collaboration*

K.C. Bowler, N.M. Hazel, H. Hoeber[1], R.D. Kenway, D.G. Richards

*Department of Physics & Astronomy, The University of Edinburgh, Edinburgh EH9 3JZ, Scotland*

L. Lellouch[2], J. Nieves, C.T. Sachrajda, H. Wittig

*Physics Department, The University, Southampton SO17 1BJ, UK*

## Abstract

We present results of a lattice computation of the matrix elements of the vector and axial-vector currents which are relevant for the semi-leptonic decays $D \to K$ and $D \to K^*$. The computations are performed in the quenched approximation to lattice QCD on a $24^3 \times 48$ lattice at $\beta = 6.2$, using an $O(a)$-improved fermionic action. In the limit of zero lepton masses the semi-leptonic decays $D \to K$ and $D \to K^*$ are described by four form factors: $f_K^+, V, A_1$ and $A_2$, which are functions of $q^2$, where $q^\mu$ is the four-momentum transferred in the process. Our results for these form factors at $q^2 = 0$ are: $f_K^+(0) = 0.67\,^{+7}_{-8}$, $V(0) = 1.01\,^{+30}_{-13}$, $A_1(0) = 0.70\,^{+7}_{-10}$, $A_2(0) = 0.66\,^{+10}_{-15}$, which are consistent with the most recent experimental world average values. We have also determined the $q^2$ dependence of the form factors, which we find to be reasonably well described by a simple pole-dominance model. Results for other form factors, including those relevant to the decays $D \to \pi$ and $D \to \rho$, are also given.

---

[1] Present address: HLRZ Jülich, D-52425 Jülich, Germany.

[2] Present address: Centre de Physique Theorique, CNRS Luminy, Case 907, F-13288 Marseille Cedex 9, France.

# 1 Introduction

Semi-leptonic decays of the heavy-light mesons have attracted considerable interest, as they play a crucial role in the determination of the elements of the Cabibbo-Kobayashi-Maskawa (CKM) mixing matrix and in the understanding of weak decays. In recent years, a machinery has been developed for calculating weak matrix elements from lattice simulations (for review lectures presented at recent lattice conferences see refs. [1]–[6]). D decays provide a good test of the method, since the relevant CKM matrix elements ($V_{cs}$ and $V_{cd}$) are well constrained in the Standard Model. In addition comparisons between D and B decays reveal the size of non-leading terms in the Heavy Quark Effective Theory (HQET).

The study of the decays $D \to K l^+ \nu_l$ and $D \to \bar{K}^* l^+ \nu_l$ (and similarly $D \to \pi l^+ \nu_l$, $D \to \rho l^+ \nu_l$) is particularly simple. They proceed via the spectator process in which a charm quark decays into a light quark ($s$ or $d$) by emitting a $W$-boson, which materializes into a lepton pair $(l^+, \nu_l)$, as shown in Figure 1. With only a single hadron in the final state, there are no interfering diagrams or final-state interactions to take into account, unlike the situation in non-leptonic decays.

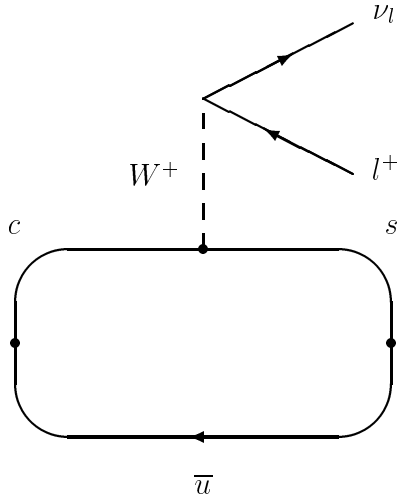

Figure 1: Feynman diagram relevant in semi-leptonic $D \to K, K^*$ decays.

The non-perturbative strong interaction effects are contained in the matrix elements $\langle K^* | J^\mu | D \rangle$ and $\langle K | J^\mu | D \rangle$, where $J^\mu = \bar{s} \gamma^\mu (1 - \gamma_5) c$ is the relevant quark weak current[3]. In this paper we present the results of a lattice calculation of these matrix elements using the improved quark action proposed by Sheikholeslami and Wohlert [7]. We determine the dependence of the form factors on the momentum transfer ($q$), and study the phenomenological implications of our results. Previous lattice studies of these decays, obtained using the Wilson quark action can be found in refs. [8]–[15], and using the Sheikholeslami-Wohlert action in ref. [16].

---

[3]This discussion applies equally well to the $D \to \pi, \rho$ cases by modifying the appropriate flavour quantum numbers.

The plan of this paper is the following. In section 2 we review the experimental situation and give the general formulae necessary for the calculation of the $D \to K l^+ \nu_l$ and $D \to \bar{K}^* l^+ \nu_l$ decay rates. In section 3 we describe the details of our simulation, and the methods used to determine the matrix elements (and hence the form factors) from the correlation functions computed on the lattice. The results, as is always the case in lattice simulations, are obtained for unphysically large values of the masses of the $u$ and $d$ quarks and have to be extrapolated to the chiral limit. Details of this extrapolation are presented in section 4, and in section 5 we discuss the relation between the lattice vector and axial currents used in this study, and the corresponding continuum currents. In section 6 we present a compendium of all our results. Finally, in section 7, we study the implications of our results, comparing them with the experimental measurements from refs. [17]–[27], summarised in [28]–[29], and with other theoretical predictions [8]–[16] and [30]–[37].

## 2 Phenomenology

Using Lorentz, parity and time-reversal invariance, the matrix elements for the decays $D \to K$ and $D \to K^*$, can be parametrized (in Minkowski space) in terms of invariant form factors as follows [9, 11, 12, 30]:

$$\langle K | (V-A)_\mu | D \rangle = \left( p_D + p_K - q \frac{m_D^2 - m_K^2}{q^2} \right)_\mu f^+(q^2) + q_\mu \frac{m_D^2 - m_K^2}{q^2} f^0(q^2) \quad (1)$$

$$\langle K_r^* | (V-A)_\mu | D \rangle = \epsilon_r^{*\beta} T_{\mu\beta} \quad (2)$$

$$T_{\mu\beta} = \frac{2V(q^2)}{m_D + m_{K^*}} \epsilon_{\mu\gamma\delta\beta} p_D^\gamma p_{K^*}^\delta - i(m_D + m_{K^*}) A_1(q^2) g_{\mu\beta}$$
$$+ i \frac{A_2(q^2)}{m_D + m_{K^*}} (p_D + p_{K^*})_\mu q_\beta - i \frac{A(q^2)}{q^2} 2 m_{K^*} q_\mu (p_D + p_{K^*})_\beta \quad (3)$$

where $q^\mu = (p_D - p_{K(K^*)})^\mu$ is the four-momentum transfer, $\epsilon_r^{*\beta}$ is the polarization vector of the $K^*$ and $f^{+,0}, V, A$ and $A_{1,2}$ are dimensionless form factors. $A$ can be written as

$$A(q^2) = A_0(q^2) - A_3(q^2) \quad (4)$$

$$A_3(q^2) = \frac{m_D + m_{K^*}}{2 m_{K^*}} A_1(q^2) - \frac{m_D - m_{K^*}}{2 m_{K^*}} A_2(q^2) \quad (5)$$

with $A_0(0) = A_3(0)$ and $f^+(0) = f^0(0)$. In the limit of zero lepton masses, the terms proportional to $f^0$ in eq. (1) and to $A$ in eq. (3) do not contribute to the total amplitude and hence to the decay rates.

The physical meaning of the different form factors is clear in the helicity basis, in which each of the form factors corresponds to a transition amplitude with definite spin-parity quantum numbers in the frame of the center of mass of the lepton pair. Pole dominance models [30] then suggest the following behaviour with $q^2$:

$$V(q^2) = \frac{V(0)}{1 - q^2/m_{1^-}^2} \quad , \quad A_0(q^2) = \frac{A_0(0)}{1 - q^2/m_{0^-}^2},$$

$$A_i(q^2) = \frac{A_i(0)}{1 - q^2/m_{1^+}^2} \quad , \quad i = 1, 2, 3 \tag{6}$$

$$f^+(q^2) = \frac{f^+(0)}{1 - q^2/m_{1^-}^2} \quad , \quad f^0(q^2) = \frac{f^0(0)}{1 - q^2/m_{0^+}^2} \tag{7}$$

where $m_{J^P}$ denotes the mass of the $\bar{s}c$ meson with spin $J$ and parity $P$. This simple picture certainly has limitations. The pole-dominated form factor would vary very rapidly with $q^2$ near the end point. Another limitation is that for $f^0, A_1, A_2$ and $A_3$, the $0^+$ and $1^+$ resonances are, in most cases, not known or only poorly established. On the lattice we can, in principle, determine form factors as functions of $q^2$. Therefore, assumptions such as pole dominance are not needed. Indeed, an important motivation for lattice computations is the opportunity to test such assumptions from first principles.

The total decay rates are given by:

$$\Gamma(D \to K l^+ \nu_l) = \frac{G_F^2 |V_{cs}|^2}{192\pi^3 m_D^3} \int_0^{(m_D - m_K)^2} dq^2 \, [\lambda(q^2)]^{\frac{3}{2}} \times |f_K^+(q^2)|^2 \tag{8}$$

$$\Gamma(D \to \bar{K}^* l^+ \nu_l) = \frac{G_F^2 |V_{cs}|^2}{192\pi^3 m_D^3} \int_0^{(m_D - m_{K^*})^2} dq^2 \, q^2 [\lambda(q^2)]^{\frac{1}{2}} \times$$
$$\times \left( |H^+(q^2)|^2 + |H^-(q^2)|^2 + |H^0(q^2)|^2 \right) \tag{9}$$

where $\lambda(q^2) = (m_D^2 + m_{K,K^*}^2 - q^2)^2 - 4m_D^2 m_{K,K^*}^2$. $H^0$ comes from the contribution of the longitudinally polarized $K^*$ and is given by [34]

$$H^0(q^2) = \frac{-1}{2m_{K^*}\sqrt{q^2}} \left\{ (m_D^2 - m_{K^*}^2 - q^2)(m_D + m_{K^*})A_1(q^2) - \frac{4m_D^2 |\vec{p}_{K^*}|^2}{m_D + m_{K^*}} A_2(q^2) \right\} \tag{10}$$

where $\vec{p}_{K^*}$ is the momentum of the $K^*$ in the D-meson rest frame. $H^\pm$ correspond to the contribution of the transverse polarizations of the vector meson and are given by [34]

$$H^\pm(q^2) = -\left\{ (m_D + m_{K^*})A_1(q^2) \mp \frac{2m_D |\vec{p}_{K^*}|}{(m_D + m_{K^*})} V(q^2) \right\} \tag{11}$$

We now briefly summarise the experimental results for semi-leptonic decays of D-mesons, basing our discussion on the review articles [28]–[29]. The largest and best measured semi-leptonic decay is $D \to K l \nu_l$. There have been several experiments ([17]–[21]) which have measured the branching ratios $B(D^0 \to K^- l^+ \nu_l)$ and $B(D^+ \to \bar{K}^0 l^+ \nu_l)$. From these experiments and the total $D^0$ and $D^+$ lifetimes, one can calculate the $D^0$ and $D^+$ semi-leptonic decay rates, which should coincide by isospin symmetry. The world average value of the semi-leptonic width quoted in [29] is $\Gamma(D \to \bar{K} l^+ \nu_l) = (7.1 \pm 0.6) \times 10^{10}$ $s^{-1}$. However, by looking only at the $D^0 \to K^- l^+ \nu_l$ channel and assuming isospin symmetry, a different average value, $\Gamma(D \to \bar{K} l^+ \nu_l) = (9.0 \pm 0.5) \times 10^{10}$ $s^{-1}$, is given in [28][4].

The shape of the form factor is a measure of the decreasing overlap of the $D$ and $K$ wave functions as $E_K$ increases. CLEO has measured this shape with the largest sample of $D^0 \to K^- l^+ \nu_l$ decays. Due to the phase space, the differential decay rate peaks at low $q^2$. The factor $|f_K^+(q^2)|^2$ increases with $q^2$, changing by a factor of about 2 over the kinematical range of the decay. A good fit to the data is obtained using eq. (7) with a pole mass $m_{1^-} = (2.00 \pm 0.12 \pm 0.18)$ GeV [17], which is in good agreement with the value of 2.1 GeV expected from the closest resonance with the proper quantum numbers, the $D_s^*$. The measured value of the pole mass $m_{1^-}$ agrees with earlier experiments but with a smaller error. $f_K^+(0)$ is obtained from the total semi-leptonic width, integrated over $q^2$, by assuming pole dominance. The average value quoted in [29] is $f_K^+(0) = 0.70 \pm 0.03$, whereas the average value quoted in [28] is $f_K^+(0) = 0.77 \pm 0.04$.

The Cabibbo-suppressed decay $D \to \pi l \nu_l$ has also been observed. Since the ratio $|V_{cd}/V_{cs}|$ is known, assuming unitarity of the CKM matrix, from the comparison of the decays $D \to \pi l \nu_l$ and $D \to K l \nu_l$ it is possible to determine the ratio $f_\pi^+(0)/f_K^+(0)$. This ratio is predicted theoretically to lie in the broad range 0.7–1.4. Mark III [20] gives a result of $f_\pi^+(0)/f_K^+(0) = 1.0 {}^{+0.6}_{-0.3} \pm 0.1$. In a recent analysis, CLEO gets a value $f_\pi^+(0)/f_K^+(0) = 1.29 \pm 0.21 \pm 0.11$, [27]. The errors in this ratio of form factors are still very large. In addition in [28], a value for the rate $\Gamma(D \to \pi l^+ \nu_l)$ of $(1.2 \pm 0.3) \times 10^{10}$ $s^{-1}$ is quoted.

There have been a number of measurements of $B(D \to K^* l^+ \nu_l)$, with both $D^0$ and $D^+$ mesons ([17],[21]–[25]). The average value of the width from these measurements is $\Gamma(D \to K^* l^+ \nu_l) = (5.1 \pm 0.5) \times 10^{10}$ $s^{-1}$ [28]. A slightly different average value, $\Gamma(D \to K^* l^+ \nu_l) = (4.5 \pm 0.5) \times 10^{10}$ $s^{-1}$, is given in ref. [29]. The experimental results for the form factors $V, A_1$ and $A_2$ are summarised in Table 1 and the results for the ratio of the decay rates of the longitudinal ($H^0$ contribution in eq. (9)) and transverse ($H^\pm$ contributions in eq. (9)) $K^*$ are presented in Table 2. The total rate is dominated by the $A_1$ form factor and the ratios of form factors are determined by fitting the angular distributions.

For the (Cabibbo-suppressed) decay $D \to \rho l \nu_l$ there only exists an upper limit for the branching fraction at 90 % confidence level $B(D \to \rho l \nu_l) < 0.37$ [20].

---

[4]In [28] the most recent measurement of the CLEO collaboration ([17]) is included, whereas it is omitted in ref. [29].

| Exp. | E-687 [22] | E-653 [26] | E-691 [23] | World Ave. [28] |
|---|---|---|---|---|
| $V/A_1$ | $1.74 \pm 0.27 \pm 0.28$ | $2.00 \pm 0.33 \pm 0.16$ | $2.0 \pm 0.6 \pm 0.3$ | $1.90 \pm 0.25$ |
| $A_2/A_1$ | $0.78 \pm 0.18 \pm 0.10$ | $0.82 \pm 0.22 \pm 0.11$ | $0.0 \pm 0.5 \pm 0.2$ | $0.74 \pm 0.15$ |

| Exp. | $A_1$ | $A_2$ | $V$ |
|---|---|---|---|
| World Ave. [28] | $0.61 \pm 0.05$ | $0.45 \pm 0.09$ | $1.16 \pm 0.16$ |

Table 1: Form Factors at $q^2 = 0$ for the $D \to K^* l^+ \nu_l$ decay.

| Exp. | E-687 [22] | E-653 [26] | E-691 [23] | World Ave. [29] |
|---|---|---|---|---|
| $\Gamma_L/\Gamma_T$ | $1.20 \pm 0.13 \pm 0.13$ | $1.18 \pm 0.18 \pm 0.08$ | $1.8\,^{+0.6}_{-0.4} \pm 0.3$ | $1.2 \pm 0.1$ |

Table 2: Ratio of the longitudinal and transverse partial widths for the $D \to K^* l^+ \nu_l$ decay.

## 3 Details of the Simulation

We work in the quenched approximation on a $24^3 \times 48$ lattice at $\beta = 6.2$, which corresponds to an inverse lattice spacing $a^{-1} = 2.73 \pm 0.05$ GeV, as determined from the string tension [38]. Other physical quantities lead to slightly different values for the lattice spacing ($a^{-1} = 2.7 \sim 3.0$ GeV) [39]. This uncertainty in the determination of the scale should be reflected in our results for dimensionful quantities. Our calculation is performed on sixty $SU(3)$ gauge field configurations [38]. The gauge configurations and quark propagators were produced on the 64-node i860 Meiko Computing Surface at the University of Edinburgh. The $SU(3)$ gauge fields were generated using the Hybrid Over-Relaxed algorithm, defined in reference [38]. The gauge configurations are separated by 2400 sweeps, beginning at configuration 16800. The quark propagators were calculated using an $O(a)$-improved action proposed by Sheikholeslami and Wohlert, which we refer to as the $SW$-action [7],

$$S_F^{SW} = S_F^W - i\frac{\kappa}{2} \sum_{x,\mu,\nu} \bar{q}(x) F_{\mu\nu}(x) \sigma_{\mu\nu} q(x), \tag{12}$$

where $S_F^W$ is the standard Wilson lattice action,

$$S_F^W = \sum_x \left\{ \bar{q}(x)q(x) - \kappa \sum_\mu \left[ \bar{q}(x)(1-\gamma_\mu)U_\mu(x)q(x+\hat{\mu}) + \bar{q}(x+\hat{\mu})(1+\gamma_\mu)U_\mu^\dagger(x)q(x) \right] \right\}. \tag{13}$$

$F_{\mu\nu}$ is a lattice definition of the field strength tensor and $\kappa$ is the hopping parameter. Periodic boundary conditions were employed in the spatial direction and anti-periodic in the temporal direction. The "improvement" is particularly important here since we are studying the propagation of quarks whose bare masses are around one third of the inverse lattice spacing.

The interpolating operators and currents which we use in this study are of the form:

$$\bar{q}_1(x)\left(1 + \frac{ra}{2}\gamma \cdot \overleftarrow{D}\right)\Gamma\left(1 - \frac{ra}{2}\gamma \cdot \overrightarrow{D}\right)q_2(x) \tag{14}$$

where $\Gamma$ is one of the 16 Dirac matrices. The matrix elements of these operators computed using the $SW$-action have no discretisation errors of $O(a)$; the leading discretisation errors are of $O(\alpha_s a)$ [40].

Our statistical errors are calculated according to the bootstrap procedure described in ref. [38], for which the quoted errors on all quantities correspond to 68% confidence limits of the distribution obtained from 1000 bootstrap samples.

We have computed light quark propagators at three values of the quark mass corresponding to $\kappa = 0.14144, 0.14226$ and $0.14262$, using an over-relaxed minimal residual algorithm with red-black preconditioning and point sources and sinks. The masses of the light pseudoscalar and vector mesons which are needed for this study were obtained in ref. [39] and are summarised in Table 3. Results extrapolated to the chiral limit (found to correspond to a hopping parameter $\kappa_{crit} = 0.14315\,^{+2}_{-2}$) and to the mass of the strange quark ($\kappa_s = 0.1419\,^{+1}_{-1}$) are also tabulated. Our $D$-mesons consist of a heavy quark (with $\kappa_c = 0.129$, where the subscript $c$ stands for "charm") and one of the light antiquarks. The heavy quark with $\kappa = 0.129$ has a mass approximately equal to that of the charm quark [41]. We use spatially-extended interpolating operators for the $D$-mesons (we use gauge-invariant Jacobi smearing on the heavy-quark field, described in detail in ref. [41]), but local operators of the form in eq. (14) for the light mesons.

Further details on the lattice calibration, fitting procedures, mass spectrum and extraction of matrix elements of local operators between the vacuum and meson states, e.g. $\langle M_{PS}|\bar{Q}\gamma_5 q|0\rangle$ can be found in references [39] and [42]. In Tables 3 and 4 we show a summary of the mass spectrum found which we will use below. In the following we only present those details of the calculation which are specific to semi-leptonic decays and cannot be found in the above references.

In order to determine the matrix elements in eqs. (1) and (2) we compute the three-point correlation functions:

$$C^\mu(\vec{p}_D, \vec{q}, t_D, t_J) = \sum_{\vec{x},\vec{y}} e^{i\vec{p}_D \cdot \vec{x}} e^{i\vec{q}\cdot\vec{y}} \langle J_D(t_D, \vec{x}) J_W^\mu(t_J, \vec{y}) J_K^\dagger(0, \vec{0})\rangle \tag{15}$$

$$C^{\mu\alpha}(\vec{p}_D, \vec{q}, t_D, t_J) = \sum_{\vec{x},\vec{y}} e^{i\vec{p}_D \cdot \vec{x}} e^{i\vec{q}\cdot\vec{y}} \langle J_D(t_D, \vec{x}) J_W^\mu(t_J, \vec{y}) J_{K^*}^{\alpha\dagger}(0, \vec{0})\rangle \tag{16}$$

where $J_D$ is a spatially-extended interpolating field for the D meson [41] and $J_W^\mu$ is an $O(a)$-improved lattice operator corresponding to the continuum weak currents $\bar{s}\gamma^\mu(1-\gamma_5)c$ or $\bar{q}\gamma^\mu(1-\gamma_5)c$ ($q = u$ or $d$). $J_K$ and $J_{K^*}^\alpha$ are local interpolating operators which can annihilate the light-light pseudoscalar and vector mesons respectively. The lattice vector and axial currents (eq. (14) with $\Gamma = \gamma^\mu$ or $\gamma^\mu \gamma^5$) are related to the continuum ones by renormalisation constants $Z_V$ and $Z_A$; we will discuss the determination of these constants in section 5. To evaluate these correlators, we use the standard source method [43]. We choose $t_D = 24$ (in lattice units) and symmetrize the correlators about that point using Euclidean time reversal [44]. The position of the light meson source is fixed at the origin and we have varied the time position of the current in the interval $t_J = 7 - 16$. In Euclidean space, provided the three points in the correlators of eqs. (15) and (16) are sufficiently separated in time ($t_J, t_D - t_J \gg 1$), the ground state contribution dominates and one finds for $t_J < t_D$ :

$$C^\mu(\vec{p}_D, \vec{q}, t_D = 24, t_J) \rightarrow \frac{Z_D(|\vec{p}_D|)}{2E_D(\vec{p}_D)} \frac{Z_K(|\vec{p}_D + \vec{q}|)}{2E_K(\vec{p}_D + \vec{q})} e^{-E_D(\vec{p}_D)t_D} e^{[E_D(\vec{p}_D) - E_K(\vec{p}_D + \vec{q})]t_J} \times$$
$$\Delta(\mu) \langle K, \vec{p}_D + \vec{q} | V^\mu | D, \vec{p}_D \rangle^* \quad (17)$$

$$C^{\mu,\alpha}(\vec{p}_D, \vec{q}, t_D = 24, t_J) \rightarrow \frac{Z_D(|\vec{p}_D|)}{2E_D(\vec{p}_D)} \frac{Z_{K^*}(|\vec{p}_D + \vec{q}|)}{2E_{K^*}(\vec{p}_D + \vec{q})} e^{-E_D(\vec{p}_D)t_D} e^{[E_D(\vec{p}_D) - E_{K^*}(\vec{p}_D + \vec{q})]t_J} \times$$
$$\Delta(\mu)\Delta(\alpha) \sum_r \epsilon_r^{\alpha*}(\vec{p}_D + \vec{q}) \langle K_r^*, \vec{p}_D + \vec{q} | [V - A]^\mu | D, \vec{p}_D \rangle^* \quad (18)$$

$$\Delta(\sigma = 0) = -1 \quad , \quad \Delta(\sigma = 1, 2, 3) = 1 \quad (19)$$

| $\kappa_{l_1}$ | $\kappa_{l_2}$ | $0^-$ meson | $1^-$ meson |
|---|---|---|---|
| 0.14144 | 0.14144 | 0.298 $^{+2}_{-2}$ | 0.395 $^{+7}_{-6}$ |
| 0.14144 | 0.14226 | 0.259 $^{+2}_{-2}$ | 0.370 $^{+6}_{-5}$ |
| 0.14144 | 0.14262 | 0.241 $^{+2}_{-3}$ | 0.360 $^{+8}_{-6}$ |
| 0.14226 | 0.14226 | 0.214 $^{+2}_{-3}$ | 0.343 $^{+9}_{-7}$ |
| 0.14226 | 0.14262 | 0.192 $^{+3}_{-3}$ | 0.331 $^{+11}_{-10}$ |
| 0.14262 | 0.14262 | 0.167 $^{+3}_{-4}$ | 0.319 $^{+14}_{-13}$ |
| $\kappa_s$=0.1419 | $\kappa_{crit}$=0.14315 | 0.181 $^{+9}_{-8}$ | 0.326 $^{+13}_{-12}$ |
| $\kappa_{crit}$=0.14315 | $\kappa_{crit}$=0.14315 | 0 | 0.290 $^{+10}_{-10}$ |

Table 3: Light-light meson masses in lattice units. For the pseudoscalar channel we fit over the time range $t = 14 - 22$. For the vector channel we fit over the time range $t = 13 - 23$.

| $\kappa_l$ | $0^-$ meson | $1^-$ meson |
|---|---|---|
| 0.14144 | 0.716 $^{+2}_{-2}$ | |
| $\kappa_s$= 0.1419 | 0.701 $^{+4}_{-4}$ | 0.732 $^{+4}_{-4}$ |
| 0.14226 | 0.692 $^{+3}_{-2}$ | |
| 0.14262 | 0.683 $^{+4}_{-3}$ | |
| $\kappa_{crit}$=0.14315 | 0.665 $^{+3}_{-3}$ | 0.697 $^{+5}_{-4}$ |

Table 4: Heavy-light meson masses, $\kappa_c = 0.129$, in lattice units. We fit over the time range $t = 11 - 22$.

$E_D$ is the energy of the D-meson and its wave-function factor, $Z_D(|\vec{p}_D|) \equiv \langle 0 | J_D(0) | D, \vec{p}_D \rangle$, is a function of the meson momentum because we use spatially-extended interpolating operators. $E_K(E_{K^*})$ is the energy of the light pseudoscalar (vector) meson and the wave-function factors $Z_K(|\vec{p}|) \equiv \langle 0 | J_K(0) | K, \vec{p} \rangle$ and $Z_{K^*}(|\vec{p}|)$ ($\langle 0 | J_{K^*}^\alpha(0) | K_r^*, \vec{p} \rangle \equiv \epsilon_r^\alpha(\vec{p}) Z_{K^*}(|\vec{p}|)$) do not depend on the momentum of the meson ($\vec{p}$) because we have used local densities. The factors $\Delta(\mu)$ in eq. (17) and $\Delta(\mu)\Delta(\alpha)$ in eq. (18) come from relating $\overline{D}$-meson matrix elements, which we obtain directly from the three-point correlation functions defined in eqs. (15) and (16), to those of the D-meson which we are interested in.

The matrix elements have been computed for two values of the momentum of the D-meson ($(12a/\pi)\vec{p}_D = (0,0,0), (1,0,0)$); and all values of the momentum transfer $\vec{q}$ for which $(12a/\pi)|\vec{q}| < 2$. In order to limit the systematic errors (and also statistical ones), we will only present results for matrix elements for which both the initial- and final-state mesons have three-momenta less than or equal to $\pi/12a$. To improve statistics, we average over all equivalent momenta and the different correlators, $C^{\mu,\alpha}$ or $C^\mu$, which lead to the same matrix element.

The wavefunction factors and energies are obtained from fits to two-point correlation func-

| $\kappa_l$ | $Z_D^2(\vec{p}=(0,0,0))$ | $Z_D^2(\vec{p}=\frac{\pi}{12a}(1,0,0))$ |
|---|---|---|
| 0.14144 | 14.5 $^{+5}_{-4}$ | 10.6 $^{+4}_{-3}$ |
| 0.14226 | 12.6 $^{+5}_{-4}$ | 9.0 $^{+3}_{-3}$ |
| 0.14262 | 12.0 $^{+6}_{-5}$ | 8.4 $^{+4}_{-3}$ |

Table 5: Pseudoscalar heavy-light meson wavefunctions (in lattice units), $\kappa_c = 0.129$. Fitting ranges are the same as those in Table 4.

tions. At large times, $t$, the Euclidean correlators $G_5$ and $G_{ij}$ behave as follows:

$$G_5(t,\vec{p}) = \sum_{\vec{x}} e^{i\vec{p}\vec{x}} \langle P_5(\vec{x},t) P_5^\dagger(\vec{0},0) \rangle$$
$$\rightarrow Z_5^2(|\vec{p}|) \frac{e^{-E_5(\vec{p})\frac{T}{2}}}{E_5(\vec{p})} \cosh\left(E_5(\vec{p})[t-\frac{T}{2}]\right) \qquad (20)$$

$$G_{ij}(t,\vec{p}) = \sum_{\vec{x}} e^{i\vec{p}\vec{x}} \langle V_i(\vec{x},t) V_j^\dagger(\vec{0},0) \rangle$$
$$\rightarrow \left(-g_{ij} + \frac{p_i p_j}{m_V^2}\right) Z_V^2(|\vec{p}|) \frac{e^{-E_V(\vec{p})\frac{T}{2}}}{E_V(\vec{p})} \cosh\left(E_V(\vec{p})[t-\frac{T}{2}]\right) \qquad (21)$$

where T is the length of the lattice in the time direction, $P_5$ and $V_i$ are the pseudoscalar density and vector current with the appropriate flavour quantum numbers, and $E_5$ and $E_V$ the energies of the mesons. For light mesons we use continuum dispersion relations, i.e., $E_{5,V}(\vec{p}) = \sqrt{m_{5,V}^2 + \vec{p}^{\,2}}$ and impose $Z_5(|\vec{p}|) = Z_5(|\vec{0}|)$ [45]. These relations are well satisfied for momentum $\pi/12a$ which is the highest one we have considered. For the D meson, as mentioned above, the wave-function factors $Z_D(\vec{p})$ depend on the momentum and it is necessary to fit the corresponding two-point correlators to the asymptotic expressions of eq. (20) not only for $\vec{p} = (0,0,0)$, but also for $\vec{p} = \pi/12a(1,0,0)$. We have constrained the energy to be $E_D(\vec{p}) = \sqrt{m_D^2 + \vec{p}^{\,2}}$ and therefore have performed only a one parameter fit in order to find $Z_D(|\vec{p}|)$. The masses we have used in our study of semi-leptonic decays of D-mesons appear in Tables 3 and 4, whereas the wave-function factors appear in Tables 5 and 6.

Having determined the $Z$'s and energies, all the factors multiplying the required matrix elements on the right hand sides of eqs. (17) and (18) are known, allowing us to determine the different form factors which appear in the matrix elements. The results presented in section 6 were obtained by fitting the different $(\mu, \alpha)$ correlators, for each combination of quark masses and each momentum channel, to their respective asymptotic forms (eqs. (17) and (18)) in the time interval $t_J = 11 - 13$. We have performed correlated fits, but we only allow for correlations between different timeslices ($t_J = 11, 12, 13$) of a given $(\mu, \alpha)$ correlator

| $\kappa_{l_1}$ | $\kappa_{l_2}$ | $Z_5^2(\vec{p}=(0,0,0))$ | $Z_V^2(\vec{p}=(0,0,0))$ |
|---|---|---|---|
| 0.14144 | 0.14144 | 0.0081 $^{+3}_{-3}$ | 0.0025 $^{+2}_{-1}$ |
| 0.14144 | 0.14226 | 0.0067 $^{+4}_{-3}$ | 0.0021 $^{+2}_{-2}$ |
| 0.14144 | 0.14262 | 0.0062 $^{+4}_{-4}$ | 0.0019 $^{+2}_{-2}$ |
| 0.14226 | 0.14226 | 0.0056 $^{+4}_{-3}$ | 0.0017 $^{+2}_{-2}$ |
| 0.14226 | 0.14262 | 0.0052 $^{+4}_{-4}$ | 0.0015 $^{+2}_{-2}$ |

Table 6: Light-light meson wave-functions (in lattice units). Fitting ranges are the same as those in Table 3.

at the same quark mass and momentum.

## 4 Chiral Extrapolation

We are interested in deriving the form factors for physical values of the charm, strange and light quark masses, for a range of values of the momentum transfer $q$. We obtain these by extrapolation from the three-point correlation functions of eqs. (15) and (16) computed with a fixed charm quark mass (corresponding to $\kappa_c = 0.129$) and for three values of the light quark mass (corresponding to $\kappa_l = 0.14144$, $0.14226$ and $0.14262$) and two values of the strange quark mass (corresponding to $\kappa_{l_s} = 0.14144$ and $0.14226$). The extrapolation to the physical values of the light and strange quark masses proceeds as follows:

i) For each set of three-momenta of the initial and final state mesons, we determine each form factor for the six combinations of light and strange quark masses. The masses are extrapolated to their physical values using:

$$m_D(\kappa_c, \kappa_l) = \bar{a}_{PS} + \frac{\bar{b}_{PS}}{2}\left(\frac{1}{\kappa_l} - \frac{1}{\kappa_{crit}}\right) \tag{22}$$

$$m_V(\kappa_1, \kappa_2) = a_V + b_V \left(\frac{1}{2\kappa_1} + \frac{1}{2\kappa_2} - \frac{1}{\kappa_{crit}}\right) \tag{23}$$

$$m_{PS}^2(\kappa_1, \kappa_2) = b_{PS}\left(\frac{1}{2\kappa_1} + \frac{1}{2\kappa_2} - \frac{1}{\kappa_{crit}}\right). \tag{24}$$

This extrapolation, together with the continuum dispersion relations, also determines the value of $q^2$ corresponding to each set of three-momenta for physical quark masses (see Tables 3 and 4).

ii) For each momentum channel we extrapolate the form factors to the physical limit, using the full covariance matrix, assuming the following dependence on the quark masses:

$$F(\kappa_{l_s}, \kappa_l) = a + b\,\frac{\Delta m_D(\kappa_c, \kappa_l)}{m_D(\kappa_c, \kappa_{crit})} + c\,\frac{\Delta m_{light}(\kappa_l, \kappa_{l_s})}{m_{light}(\kappa_{crit}, \kappa_s)} + d\left(\frac{\Delta m_{light}(\kappa_l, \kappa_{l_s})}{m_{light}(\kappa_{crit}, \kappa_s)}\right)^2 \quad (25)$$

where $\Delta m_D(\kappa_c, \kappa_l)$ and $\Delta m_{light}(\kappa_l, \kappa_{l_s})$ (light stands for light pseudoscalar and light vector mesons) are defined by[5]

$$\Delta m_D(\kappa_c, \kappa_l) = m_D(\kappa_c, \kappa_l) - m_D(\kappa_c, \kappa_{crit}) \quad (26)$$

$$\Delta m_{PS}(\kappa_l, \kappa_{l_s}) = m_{PS}(\kappa_l, \kappa_{l_s}) - m_{PS}(\kappa_{crit}, \kappa_s) \quad (27)$$

$$\Delta m_V(\kappa_l, \kappa_{l_s}) = m_V(\kappa_l, \kappa_{l_s}) - m_V(\kappa_{crit}, \kappa_s) \quad (28)$$

In the decay into vector mesons, we have not kept the quadratic term $(\frac{\Delta m_{light}(\kappa_l, \kappa_{l_s})}{m_{light}(\kappa_{crit}, \kappa_s)})^2$ (from Table 3 it can be seen that, unlike in the case of the light pseudoscalar meson, this term is always smaller than 5% and has a negligible effect on the extrapolation of the form factors to the physical limit) and we end up with only three free parameters $(a, b, c)$. Thus, in the $0^- \to 0^-$ case, we fit the form factors to the following dependence on the quark masses:

$$F(\kappa_{l_s}, \kappa_l) = \alpha + \beta\left(\frac{1}{\kappa_l} - \frac{1}{\kappa_{crit}}\right) + \gamma\left(\frac{1}{\kappa_{l_s}} + \frac{1}{\kappa_l} - \frac{2}{\kappa_{crit}}\right)^{\frac{1}{2}} + \delta\left(\frac{1}{\kappa_{l_s}} + \frac{1}{\kappa_l} - \frac{2}{\kappa_{crit}}\right) \quad (29)$$

and in the $0^- \to 1^-$ case we have assumed the following dependence:

$$F(\kappa_{l_s}, \kappa_l) = \alpha' + \beta'\left(\frac{1}{\kappa_l} - \frac{1}{\kappa_{crit}}\right) + \gamma'\left(\frac{1}{\kappa_{l_s}} + \frac{1}{\kappa_l} - \frac{2}{\kappa_{crit}}\right) \quad (30)$$

where $F$ represents a generic form factor. Note that, in contrast to some analyses (e.g. [13, 16]), we do not assume flavour symmetry between the active and spectator light quarks. Thus for example, the form factors extrapolated to the strange and critical quark masses, $F^K$ and $F^\rho$, are:

$$F^K = \alpha + \gamma\left(\frac{1}{\kappa_s} - \frac{1}{\kappa_{crit}}\right)^{\frac{1}{2}} + \delta\left(\frac{1}{\kappa_s} - \frac{1}{\kappa_{crit}}\right) \quad (31)$$

$$F^\rho = \alpha' \quad (32)$$

---

[5]The dependence assumed in eq. (25) is motivated by the results of a Taylor expansion of $q^2(\kappa_l, \kappa_{l_s})$ around $q^2(\kappa_{crit}, \kappa_s)$.

# 5 Renormalisation Constants $Z_V$ and $Z_A$

In this section we discuss the difficulties in determining the form factors of semi-leptonic $D \to K, K^*$ decays, due to the presence of discretisation errors. Of course these errors are substantially reduced by the use of the improved action ([7], [40]), nevertheless even in this case we believe that, for currently accessible values of the lattice spacing, they lead to uncertainties of the order of 10% in the form factors. We will now attempt to justify this statement.

The lattice currents $V_\mu^L$ and $A_\mu^L$ used in this study are related to the physical ones ($V_\mu$ and $A_\mu$) by renormalisation constants $Z_V$ and $Z_A$:

$$Z_V V_\mu^L = V_\mu \quad ; \quad Z_A A_\mu^L = A_\mu \tag{33}$$

When both quarks are light, at $\beta = 6.2$, $Z_V$ and $Z_A$ are known to be about 0.83 and 1.05 respectively [46]. For light quark masses the discretisation errors are very small, but for the charmed quark at $\beta = 6.2$ this is no longer the case. In previous simulations, using Wilson fermions, these effects were modelled by using an effective (mass–dependent) value of $Z_V$ and $Z_A$, or by assuming that the "conserved" vector current, i.e. the lattice current which would be conserved if the quarks were degenerate, is free of discretisation errors, [45]. However we wish to stress that the discretisation errors of $O(\alpha_s m a)$ and $O(m^2 a^2)$ are in general different for matrix elements of currents with different Lorentz indices and between different states. Thus they cannot be absorbed into an effective $Z_V$ or $Z_A$ for all cases. To see this, note that there are discretisation errors due to the mixing of the currents with higher dimensional operators, e.g. the vector current can mix with $a\bar{q}_1 D^\mu q_2$ or $a^2 \bar{q}_1 \gamma^\mu D^\mu D^\mu q_2$. The behaviour of matrix elements of these operators with the external states is in general different from that of the currents. We have carried out an extensive study of these effects for the heavy-heavy vector current $\bar{Q}\gamma^\mu Q$[6].

Defining $Z_V^{eff}$ by

$$Z_V^{eff} \equiv \frac{C_2(t_x; \vec{p})}{C_3^\mu(t_y, t_x; \vec{p})} \frac{p^\mu}{E} \tag{34}$$

where

$$C_2(t_x; \vec{p}) = \sum_{\vec{x}} e^{i\vec{p}\cdot\vec{x}} \langle J_P(x) J_P^\dagger(0) \rangle \tag{35}$$

$$C_3^\mu(t_y, t_x; \vec{p}) = \sum_{\vec{x},\vec{y}} e^{i\vec{p}\cdot\vec{x}} \langle J_P(x) V^\mu(y) J_P^\dagger(0) \rangle \tag{36}$$

and $J_P^\dagger$ and $J_P$ are the interpolating operators which can create or annihilate the heavy-light pseudoscalar meson P. For degenerate quarks with $\kappa = 0.129$, and using correlation functions with $\vec{p} = \vec{0}$ and $\mu = 4$, we find $Z_V^{eff} = 0.9177 \, ^{+3}_{-2}$. We note that this value differs by about 10% from that for $Z_V$ determined using light quarks and is a measure of the residual size of the discretisation errors using the improved action. Another important question is whether the effects are multiplicative. To study this we have computed $Z_V^{eff}$ for $|\vec{p}_i| = |\vec{p}_f| = \frac{\pi}{12a}$, and using the current $V^4$ in the correlation function $C_3$ we find $Z_V^{eff} = 0.925(1)$ whereas using the current $V^1$ we find $Z_V^{eff} = 0.99(6)$. Unfortunately the latter error is too large for us to be able to determine whether the discretisation errors are the same in matrix elements[7] of $V^4$ and $V^1$. In any case, because of hypercubic group invariance, there are no discretisation errors of order $O(a)$ which affect spatial components of the currents in a different way to temporal components at each value of $q^2$; the leading discretisation errors for which this happens are of order $O(a^2)$ for the SW action (arising e.g. from matrix elements of the operator $a^2 \bar{q}_1 \gamma^\mu D^\mu D^\mu q_2$).

---

[6]Full details of this study can be found in [47].

For simulations using the Wilson fermion action, the discretisation errors are much larger. Kronfeld and Mackenzie ([49]) have argued that much of this uncertainty can be absorbed into a multiplicative $m$-dependent correction factor to the heavy quark propagator. Consider the continuum free propagator at zero three-momentum:

$$\int d^3 x S(x, 0) = \frac{1 + \gamma_0}{2} e^{-mt} \tag{37}$$

Using Wilson fermions the analogous expression is

$$\int d^3 x S^W(x, 0) = \frac{1 + \gamma_0}{2}(1 + m_0 a)^{-\frac{t+a}{a}} \tag{38}$$
$$= \frac{1 + \gamma_0}{2} e^{-mt} e^{-ma} \tag{39}$$

where $m = \log(1 + m_0 a)/a$ and $m_0$ is the bare mass. An important correction proposed by Kronfeld and Mackenzie is the $e^{-ma}$ factor in eq. (39). With the SW action, even at the tree level, there can be no $O(ma)$ term in the correction factor and we find

$$\int d^3 x S^{SW}(x, 0) = \frac{1 + \gamma_0}{2} \left(1 + \frac{1}{4}(1 + m_0 a - \frac{1}{1 + m_0 a})\right)^2 (1 + m_0 a)^{-\frac{t+a}{a}} \tag{40}$$

Numerically we estimate that this correction factor differs from 1 by only about 1.5%, for our charmed quark ($\kappa = 0.129$), whereas we have seen that $Z_V^{eff}$ in heavy-to-heavy transitions is about 10% larger than $Z_V$. This suggests that this correction factor accounts for only a modest part of the discretisation errors. This conclusion is supported by the behaviour of $Z_V^{eff}$ with mass, for which we find that the quadratic term (i.e. the $O(m^2 a^2)$ term) is relatively small [47].

Our conclusions from the above observations are as follows:

---

[7] There is some mild evidence [47] that $Z_V^{eff}$ may grow faster with mass for $\mu = 1$ than for $\mu = 4$ in heavy-to-heavy transitions.

i) The use of the free quark propagator is not a useful guide to the discretisation errors when using the SW action. For the Wilson action it is possible that the corresponding factor $e^{-ma}$ accounts for part of the errors. However, the remaining uncertainties are not understood, and are in any case formally at least as large as for the SW-action (i.e. there are $O(\alpha_s ma)$ terms, and there is even no proof that there are no $O(ma)$ terms above tree level[8]).

ii) Formally there is no reason to believe that the discretisation errors can be absorbed into universal effective renormalisation constants $Z_V^{eff}$ and $Z_A^{eff}$. Even if $O(m^2 a^2)$ corrections are neglected, in which case the discretisation errors are independent of the Lorentz index of the current, these discretisation errors could be different for different form factors and they could have a different $q^2$ dependence than the form factors themselves. However, if for a given form factor, both the form factor itself and the discretisation errors have the same $q^2$ dependence, e.g. if the pole dominance formula is a good approximation to both, then the corresponding effective renormalisation constant, $Z_V^{eff}$ or $Z_A^{eff}$, is independent of $q^2$ (up to corrections of $O(m^2 a^2)$). Therefore if $O(m^2 a^2)$ corrections are neglected and assuming that the meson pole dominance model describes well the $q^2$ dependence of the different form factors and their discretisation errors, only the fact that discretisation errors could, in general, be different for different form factors, prevents the absorption of all of them into universal effective renormalisation constants. The situation is not improved by the use of "conserved" currents on the lattice. Here the situation is more difficult than in the evaluation of the Isgur-Wise function, where one only needs to evaluate a single form factor and this factorization is still possible.

Given this discussion what can be done?

In this study we recognise that discretisation errors are of $O(10\%)$, and in spite of the discussion above, we assume that they can be modelled by $Z_V^{eff}$ and $Z_A^{eff}$ (at least part of the errors can be so absorbed). Specifically for the vector current we take

$$Z_V^{eff} = 0.88 \, ^{+\,4}_{-\,5} \qquad (41)$$

which represents an approximate average of $Z_V = 0.83$ (obtained with light quarks) and $Z_V^{eff} = 0.92$ for the heavy-heavy current with the mass of the heavy quark corresponding to $\kappa = 0.129$. For the axial current, a non-perturbative determination of $Z_A$ (when both quarks are light[9]) using a method based on Ward Identities [48], gives a value of $Z_A = 1.05(1)$, [46]. A one-loop calculation in perturbation theory for the SW action ([50]) when the "boosted" coupling suggested in [51] is used, gives $Z_A = 0.97$. Unlike the case of the vector current, perturbative and non-perturbative determinations of $Z_A$ do not agree for the light-light current, thus for the axial current we have decided to take

$$Z_A^{eff} = 1.05 \ ^{+\ 1}_{-\ 8} \qquad (42)$$

which corresponds to the non-perturbative determination of [46], but with an increased lower error in order to account for the perturbative value mentioned above.

---

[8]Note that in $n$th order perturbation theory, in general, terms appear which behave as $\alpha^n ma \log^n(a) \approx ma$ [40].

[9]In this case we do not have a non perturbative determination of $Z_A^{eff}$ for the heavy-heavy current.

Below we will discuss briefly the dependence of the results on the values in eqs. (41) and (42).

Over the next few years, as high statistics simulations are performed at different values of $\beta$, it will become possible to study the discretisation errors in detail.

# 6 Results

In this section we present our results for the form factors obtained at six combinations of momenta of the initial and final state mesons, which are (in units of $\pi/12a$, and using the notation $\vec{p}_D \to \vec{p}_{K,K^*}$): a) $(0,0,0) \to (0,0,0)$, b) $(0,0,0) \to (1,0,0)$, c) $(1,0,0) \to (1,0,0)$, d) $(1,0,0) \to (0,0,0)$, e) $(1,0,0) \to (-1,0,0)$, and f) $(1,0,0) \to (0,1,0)$. The momenta of the initial state $D$-meson are fixed to be $(0,0,0)$ or $(1,0,0)$, but we average over all equivalent momenta of the light meson, so that, for example, case f) is really the average of the four terms in which the final state meson has momentum $\pi/12a$ in the positive or negative $y$ or $z$ directions.

The results for the form factors, together with the corresponding values of $q^2$, are presented in Tables 7, 8 and 9. In these tables we also present the form factors extrapolated to physical quark masses for the decays $D \to K, K^*$, following the procedure described in section 4.

From the results of Tables 7, 8 and 9, one can in principle check the pole dominance relations given in eqs. (6) and (7). This is true in practice for some of the form factors. However, we have a very poor determination of both, the scalar ($0^+$) and axial-vector ($1^+$) meson masses. Thus, we have decided to extract both the pole masses ($m_{J^P}$) and the form factors at $q^2 = 0$, by fitting the chirally–extrapolated data to the pole dominance model. In the case of $f^+$, $V$ and $A_0$ we will compare the masses of the $0^-$ and $1^-$ mesons obtained from the pole dominance fit with the lattice results obtained in our simulation (Table 4).

## 6.1 The exclusive $0^- \to 0^-$ case:

With the method described in the former sections we have found for the $D \to K$ decay,

| $\vec{p}_D$ | $\vec{p}_K$ | $\kappa_l$ | $\kappa_{l_s}$ | $q^2 a^2$ | $f^+(q^2)/Z_V^{eff}$ | $f^0(q^2)/Z_V^{eff}$ |
|---|---|---|---|---|---|---|
| (0,0,0) | (0,0,0) | 0.14144 | 0.14144 | 0.175 $^{+2}_{-2}$ | | 1.03 $^{+4}_{-4}$ |
| | | | 0.14226 | 0.209 $^{+2}_{-2}$ | | 0.99 $^{+5}_{-5}$ |
| | | 0.14226 | 0.14144 | 0.188 $^{+2}_{-2}$ | | 1.04 $^{+5}_{-5}$ |
| | | | 0.14226 | 0.229 $^{+3}_{-3}$ | | 1.00 $^{+6}_{-6}$ |
| | | 0.14262 | 0.14144 | 0.195 $^{+3}_{-3}$ | | 1.05 $^{+6}_{-6}$ |
| | | | 0.14226 | 0.241 $^{+4}_{-3}$ | | 1.03 $^{+8}_{-7}$ |
| | | $\kappa_{crit}$ | $\kappa_s$ | 0.235 $^{+8}_{-9}$ | | 1.01 $^{+8}_{-7}$ |
| (0,0,0) | (1,0,0) | 0.14144 | 0.14144 | 0.033 $^{+1}_{-1}$ | 0.91 $^{+3}_{-3}$ | 0.87 $^{+3}_{-3}$ |
| | | | 0.14226 | 0.052 $^{+1}_{-1}$ | 0.88 $^{+4}_{-3}$ | 0.81 $^{+3}_{-3}$ |
| | | 0.14226 | 0.14144 | 0.037 $^{+2}_{-2}$ | 0.90 $^{+4}_{-4}$ | 0.85 $^{+3}_{-3}$ |
| | | | 0.14226 | 0.057 $^{+2}_{-2}$ | 0.86 $^{+5}_{-4}$ | 0.78 $^{+4}_{-3}$ |
| | | 0.14262 | 0.14144 | 0.039 $^{+2}_{-2}$ | 0.89 $^{+4}_{-4}$ | 0.83 $^{+4}_{-4}$ |
| | | | 0.14226 | 0.060 $^{+3}_{-2}$ | 0.84 $^{+6}_{-5}$ | 0.75 $^{+5}_{-4}$ |
| | | $\kappa_{crit}$ | $\kappa_s$ | 0.052 $^{+4}_{-4}$ | 0.88 $^{+5}_{-5}$ | 0.80 $^{+4}_{-4}$ |
| (1,0,0) | (1,0,0) | 0.14144 | 0.14144 | 0.134 $^{+1}_{-1}$ | 1.13 $^{+7}_{-7}$ | 0.91 $^{+6}_{-7}$ |
| | | | 0.14226 | 0.155 $^{+2}_{-1}$ | 1.12 $^{+8}_{-8}$ | 0.82 $^{+7}_{-7}$ |
| | | 0.14226 | 0.14144 | 0.138 $^{+2}_{-2}$ | 1.09 $^{+11}_{-11}$ | 0.77 $^{+10}_{-11}$ |
| | | | 0.14226 | 0.162 $^{+2}_{-2}$ | 1.07 $^{+13}_{-14}$ | 0.64 $^{+11}_{-11}$ |
| | | 0.14262 | 0.14144 | 0.141 $^{+2}_{-2}$ | 1.00 $^{+14}_{-15}$ | 0.58 $^{+14}_{-15}$ |
| | | | 0.14226 | 0.165 $^{+3}_{-2}$ | 0.94 $^{+19}_{-21}$ | 0.40 $^{+16}_{-17}$ |
| | | $\kappa_{crit}$ | $\kappa_s$ | 0.157 $^{+4}_{-4}$ | 1.16 $^{+20}_{-19}$ | 0.73 $^{+15}_{-15}$ |

Table 7: Form factors for $0^- \to 0^-$ decay with momenta in units of $\frac{\pi}{12a}$ and $\kappa_c = 0.129$. For $f^0(q^2)$, the channel $(1,0,0) \to (1,0,0)$ has not been considered in the pole dominance fit, because we feel we do not control its chiral extrapolation.

| $\vec{p}_D$ | $\vec{p}_K$ | $\kappa_l$ | $\kappa_{l_s}$ | $q^2 a^2$ | $f^+(q^2)/Z_V^{eff}$ | $f^0(q^2)/Z_V^{eff}$ |
|---|---|---|---|---|---|---|
| (1,0,0) | (0,0,0) | 0.14144 | 0.14144 | 0.147 $^{+2}_{-2}$ | 1.25 $^{+5}_{-5}$ | 0.98 $^{+4}_{-4}$ |
| | | | 0.14226 | 0.185 $^{+2}_{-2}$ | 1.33 $^{+7}_{-7}$ | 0.96 $^{+4}_{-4}$ |
| | | 0.14226 | 0.14144 | 0.163 $^{+3}_{-2}$ | 1.27 $^{+7}_{-7}$ | 0.99 $^{+5}_{-4}$ |
| | | | 0.14226 | 0.208 $^{+3}_{-3}$ | 1.40 $^{+10}_{-10}$ | 1.00 $^{+6}_{-5}$ |
| | | 0.14262 | 0.14144 | 0.172 $^{+3}_{-3}$ | 1.26 $^{+9}_{-10}$ | 1.00 $^{+5}_{-5}$ |
| | | | 0.14226 | 0.223 $^{+4}_{-3}$ | 1.43 $^{+14}_{-15}$ | 1.01 $^{+6}_{-7}$ |
| | | $\kappa_{crit}$ | $\kappa_s$ | 0.217 $^{+9}_{-9}$ | 1.48 $^{+16}_{-17}$ | 1.06 $^{+7}_{-8}$ |
| (1,0,0) | (-1,0,0) | 0.14144 | 0.14144 | $-$ 0.140 $^{+1}_{-1}$ | 0.60 $^{+3}_{-3}$ | 0.71 $^{+4}_{-3}$ |
| | | | 0.14226 | $-$ 0.119 $^{+2}_{-1}$ | 0.58 $^{+3}_{-3}$ | 0.69 $^{+4}_{-4}$ |
| | | 0.14226 | 0.14144 | $-$ 0.136 $^{+2}_{-2}$ | 0.59 $^{+5}_{-5}$ | 0.71 $^{+6}_{-5}$ |
| | | | 0.14226 | $-$ 0.113 $^{+2}_{-2}$ | 0.58 $^{+6}_{-5}$ | 0.69 $^{+7}_{-6}$ |
| | | 0.14262 | 0.14144 | $-$ 0.133 $^{+2}_{-2}$ | 0.58 $^{+7}_{-7}$ | 0.69 $^{+9}_{-8}$ |
| | | | 0.14226 | $-$ 0.109 $^{+3}_{-2}$ | 0.57 $^{+10}_{-8}$ | 0.67 $^{+11}_{-10}$ |
| | | $\kappa_{crit}$ | $\kappa_s$ | $-$ 0.117 $^{+4}_{-4}$ | 0.63 $^{+8}_{-8}$ | 0.73 $^{+9}_{-9}$ |
| (1,0,0) | (0,1,0) | 0.14144 | 0.14144 | $-$ 0.003 $^{+1}_{-1}$ | 0.78 $^{+3}_{-3}$ | 0.79 $^{+3}_{-3}$ |
| | | | 0.14226 | 0.018 $^{+2}_{-1}$ | 0.75 $^{+4}_{-4}$ | 0.73 $^{+4}_{-4}$ |
| | | 0.14226 | 0.14144 | 0.001 $^{+2}_{-2}$ | 0.76 $^{+4}_{-4}$ | 0.76 $^{+4}_{-4}$ |
| | | | 0.14226 | 0.025 $^{+2}_{-2}$ | 0.73 $^{+5}_{-5}$ | 0.70 $^{+5}_{-5}$ |
| | | 0.14262 | 0.14144 | 0.004 $^{+2}_{-2}$ | 0.74 $^{+6}_{-6}$ | 0.74 $^{+5}_{-5}$ |
| | | | 0.14226 | 0.028 $^{+3}_{-2}$ | 0.71 $^{+7}_{-7}$ | 0.67 $^{+7}_{-7}$ |
| | | $\kappa_{crit}$ | $\kappa_s$ | 0.020 $^{+4}_{-4}$ | 0.75 $^{+5}_{-6}$ | 0.72 $^{+5}_{-5}$ |

Table 7: (cont.)

| $\vec{p}_D$ | $\vec{p}_{K^*}$ | $\kappa_l$ | $\kappa_{l_s}$ | $q^2 a^2$ | $V(q^2)/Z_V^{eff}$ |
|---|---|---|---|---|---|
| (0,0,0) | (0,0,0) | 0.14144 | 0.14144 | 0.103 $^{+2}_{-2}$ | |
| | | | 0.14226 | 0.119 $^{+3}_{-3}$ | |
| | | 0.14226 | 0.14144 | 0.103 $^{+3}_{-3}$ | |
| | | | 0.14226 | 0.121 $^{+5}_{-5}$ | |
| | | 0.14262 | 0.14144 | 0.104 $^{+4}_{-4}$ | |
| | | | 0.14226 | 0.124 $^{+7}_{-6}$ | |
| | | $\kappa_{crit}$ | $\kappa_s$ | 0.112 $^{+5}_{-5}$ | |
| (0,0,0) | (1,0,0) | 0.14144 | 0.14144 | $-$0.010 $^{+2}_{-1}$ | 1.50 $^{+5}_{-5}$ |
| | | | 0.14226 | 0.000 $^{+2}_{-2}$ | 1.47 $^{+7}_{-7}$ |
| | | 0.14226 | 0.14144 | $-$0.012 $^{+2}_{-2}$ | 1.45 $^{+8}_{-9}$ |
| | | | 0.14226 | $-$0.001 $^{+3}_{-3}$ | 1.41 $^{+10}_{-11}$ |
| | | 0.14262 | 0.14144 | $-$0.012 $^{+3}_{-2}$ | 1.43 $^{+12}_{-12}$ |
| | | | 0.14226 | $-$0.001 $^{+4}_{-4}$ | 1.36 $^{+16}_{-15}$ |
| | | $\kappa_{crit}$ | $\kappa_s$ | $-$0.009 $^{+3}_{-3}$ | 1.40 $^{+13}_{-12}$ |
| (1,0,0) | (1,0,0) | 0.14144 | 0.14144 | 0.083 $^{+2}_{-2}$ | 2.7 $^{+4}_{-4}$ |
| | | | 0.14226 | 0.095 $^{+3}_{-2}$ | 2.9 $^{+5}_{-6}$ |
| | | 0.14226 | 0.14144 | 0.082 $^{+3}_{-2}$ | 3.4 $^{+7}_{-7}$ |
| | | | 0.14226 | 0.095 $^{+4}_{-3}$ | 4.1 $^{+9}_{-9}$ |
| | | 0.14262 | 0.14144 | 0.082 $^{+3}_{-3}$ | 3.7 $^{+10}_{-10}$ |
| | | | 0.14226 | 0.096 $^{+5}_{-5}$ | 5.2 $^{+13}_{-13}$ |
| | | $\kappa_{crit}$ | $\kappa_s$ | 0.086 $^{+3}_{-4}$ | 3.7 $^{+8}_{-7}$ |

Table 8: Vector form factor ($V(q^2)$) for $0^- \to 1^-$ decay with momenta in units of $\frac{\pi}{12a}$ and $\kappa_c = 0.129$. The channel $(1,0,0) \to (1,0,0)$ has not been considered in the pole dominance fit, because we feel we do not control its chiral extrapolation.

| $\vec{p}_D$ | $\vec{p}_{K^*}$ | $\kappa_l$ | $\kappa_{l_s}$ | $q^2 a^2$ | $V(q^2)/Z_V^{eff}$ |
|---|---|---|---|---|---|
| (1,0,0) | (0,0,0) | 0.14144 | 0.14144 | $0.066\ ^{+3}_{-3}$ | $1.46\ ^{+10}_{-10}$ |
| | | | 0.14226 | $0.085\ ^{+4}_{-4}$ | $1.39\ ^{+14}_{-14}$ |
| | | 0.14226 | 0.14144 | $0.068\ ^{+4}_{-3}$ | $1.36\ ^{+16}_{-18}$ |
| | | | 0.14226 | $0.088\ ^{+6}_{-5}$ | $1.30\ ^{+25}_{-26}$ |
| | | 0.14262 | 0.14144 | $0.069\ ^{+5}_{-4}$ | $1.42\ ^{+28}_{-31}$ |
| | | | 0.14226 | $0.092\ ^{+8}_{-7}$ | $1.42\ ^{+38}_{-42}$ |
| | | $\kappa_{crit}$ | $\kappa_s$ | $0.079\ ^{+5}_{-6}$ | $1.09\ ^{+16}_{-16}$ |
| (1,0,0) | (-1,0,0) | 0.14144 | 0.14144 | $-0.191\ ^{+2}_{-2}$ | $0.90\ ^{+6}_{-7}$ |
| | | | 0.14226 | $-0.179\ ^{+3}_{-2}$ | $0.90\ ^{+9}_{-9}$ |
| | | 0.14226 | 0.14144 | $-0.192\ ^{+3}_{-2}$ | $0.85\ ^{+11}_{-12}$ |
| | | | 0.14226 | $-0.179\ ^{+4}_{-3}$ | $0.87\ ^{+15}_{-15}$ |
| | | 0.14262 | 0.14144 | $-0.192\ ^{+3}_{-3}$ | $0.81\ ^{+18}_{-18}$ |
| | | | 0.14226 | $-0.179\ ^{+5}_{-5}$ | $0.81\ ^{+24}_{-25}$ |
| | | $\kappa_{crit}$ | $\kappa_s$ | $-0.188\ ^{+3}_{-4}$ | $0.80\ ^{+12}_{-13}$ |
| (1,0,0) | (0,1,0) | 0.14144 | 0.14144 | $-0.054\ ^{+2}_{-2}$ | $1.22\ ^{+6}_{-6}$ |
| | | | 0.14226 | $-0.042\ ^{+3}_{-2}$ | $1.18\ ^{+7}_{-8}$ |
| | | 0.14226 | 0.14144 | $-0.055\ ^{+3}_{-2}$ | $1.11\ ^{+9}_{-11}$ |
| | | | 0.14226 | $-0.042\ ^{+4}_{-3}$ | $1.06\ ^{+11}_{-14}$ |
| | | 0.14262 | 0.14144 | $-0.055\ ^{+3}_{-3}$ | $1.00\ ^{+16}_{-20}$ |
| | | | 0.14226 | $-0.042\ ^{+5}_{-5}$ | $0.90\ ^{+20}_{-24}$ |
| | | $\kappa_{crit}$ | $\kappa_s$ | $-0.051\ ^{+3}_{-4}$ | $1.02\ ^{+12}_{-14}$ |

Table 8: (cont.)

| $\vec{p}_D$ | $\vec{p}_K$ | $\kappa_l$ | $\kappa_{l_s}$ | $A_1(q^2)/Z_A^{eff}$ | $A_2(q^2)/Z_A^{eff}$ | $A_0(q^2)/Z_A^{eff}$ |
|---|---|---|---|---|---|---|
| (0,0,0) | (0,0,0) | 0.14144 | 0.14144 | 0.74 $^{+2}_{-2}$ | | |
| | | | 0.14226 | 0.70 $^{+3}_{-2}$ | | |
| | | 0.14226 | 0.14144 | 0.76 $^{+3}_{-3}$ | | |
| | | | 0.14226 | 0.72 $^{+4}_{-4}$ | | |
| | | 0.14262 | 0.14144 | 0.78 $^{+5}_{-5}$ | | |
| | | | 0.14226 | 0.73 $^{+6}_{-6}$ | | |
| | | $\kappa_{crit}$ | $\kappa_s$ | 0.76 $^{+4}_{-4}$ | | |
| (0,0,0) | (1,0,0) | 0.14144 | 0.14144 | 0.70 $^{+2}_{-2}$ | 0.59 $^{+6}_{-7}$ | 0.73 $^{+2}_{-2}$ |
| | | | 0.14226 | 0.65 $^{+2}_{-2}$ | 0.51 $^{+7}_{-7}$ | 0.72 $^{+3}_{-3}$ |
| | | 0.14226 | 0.14144 | 0.71 $^{+3}_{-4}$ | 0.63 $^{+9}_{-9}$ | 0.73 $^{+3}_{-3}$ |
| | | | 0.14226 | 0.66 $^{+4}_{-4}$ | 0.53 $^{+11}_{-12}$ | 0.72 $^{+4}_{-4}$ |
| | | 0.14262 | 0.14144 | 0.71 $^{+5}_{-5}$ | 0.61 $^{+13}_{-14}$ | 0.74 $^{+5}_{-5}$ |
| | | | 0.14226 | 0.64 $^{+6}_{-6}$ | 0.49 $^{+17}_{-18}$ | 0.72 $^{+6}_{-5}$ |
| | | $\kappa_{crit}$ | $\kappa_s$ | 0.72 $^{+4}_{-4}$ | 0.69 $^{+10}_{-10}$ | 0.72 $^{+4}_{-4}$ |
| (1,0,0) | (1,0,0) | 0.14144 | 0.14144 | 0.76 $^{+7}_{-7}$ | 1.4 $^{+11}_{-11}$ | 0.92 $^{+11}_{-11}$ |
| | | | 0.14226 | 0.71 $^{+9}_{-9}$ | 1.5 $^{+12}_{-12}$ | 0.96 $^{+15}_{-15}$ |
| | | 0.14226 | 0.14144 | 0.66 $^{+11}_{-12}$ | 0.7 $^{+17}_{-17}$ | 0.74 $^{+18}_{-19}$ |
| | | | 0.14226 | 0.60 $^{+15}_{-16}$ | 0.7 $^{+20}_{-20}$ | 0.78 $^{+25}_{-25}$ |
| | | 0.14262 | 0.14144 | 0.49 $^{+19}_{-21}$ | $-0.7$ $^{+26}_{-26}$ | 0.53 $^{+27}_{-30}$ |
| | | | 0.14226 | 0.41 $^{+26}_{-27}$ | $-0.9$ $^{+30}_{-30}$ | 0.57 $^{+36}_{-38}$ |
| | | $\kappa_{crit}$ | $\kappa_s$ | 0.68 $^{+14}_{-15}$ | 1.4 $^{+23}_{-23}$ | 0.78 $^{+23}_{-24}$ |

Table 9: Axial form factors for $0^- \to 1^-$ decay with momenta in units of $\frac{\pi}{12a}$ and $\kappa_c = 0.129$. For $A_2$, the channel $(1,0,0) \to (1,0,0)$ has not been considered in the pole dominance fit, because we feel we do not control its chiral extrapolation.

| $\vec{p}_D$ | $\vec{p}_K$ | $\kappa_l$ | $\kappa_{l_s}$ | $A_1(q^2)/Z_A^{eff}$ | $A_2(q^2)/Z_A^{eff}$ | $A_0(q^2)/Z_A^{eff}$ |
|---|---|---|---|---|---|---|
| (1,0,0) | (0,0,0) | 0.14144 | 0.14144 | 0.69 $^{+3}_{-3}$ | 0.67 $^{+39}_{-38}$ | 0.88 $^{+6}_{-7}$ |
| | | | 0.14226 | 0.66 $^{+3}_{-3}$ | 0.93 $^{+50}_{-46}$ | 0.88 $^{+9}_{-9}$ |
| | | 0.14226 | 0.14144 | 0.69 $^{+4}_{-4}$ | 0.57 $^{+55}_{-52}$ | 0.92 $^{+9}_{-10}$ |
| | | | 0.14226 | 0.66 $^{+5}_{-4}$ | 0.92 $^{+78}_{-71}$ | 0.91 $^{+13}_{-14}$ |
| | | 0.14262 | 0.14144 | 0.69 $^{+6}_{-6}$ | 0.37 $^{+75}_{-71}$ | 0.99 $^{+12}_{-14}$ |
| | | | 0.14226 | 0.65 $^{+7}_{-6}$ | 0.76 $^{+100}_{-100}$ | 0.99 $^{+18}_{-21}$ |
| | | $\kappa_{crit}$ | $\kappa_s$ | 0.66 $^{+5}_{-5}$ | 0.51 $^{+75}_{-66}$ | 0.90 $^{+12}_{-14}$ |
| (1,0,0) | (-1,0,0) | 0.14144 | 0.14144 | 0.57 $^{+4}_{-4}$ | 0.36 $^{+8}_{-7}$ | 0.44 $^{+3}_{-3}$ |
| | | | 0.14226 | 0.53 $^{+5}_{-5}$ | 0.30 $^{+9}_{-8}$ | 0.44 $^{+3}_{-4}$ |
| | | 0.14226 | 0.14144 | 0.59 $^{+7}_{-8}$ | 0.37 $^{+12}_{-12}$ | 0.45 $^{+4}_{-5}$ |
| | | | 0.14226 | 0.56 $^{+9}_{-9}$ | 0.31 $^{+15}_{-15}$ | 0.47 $^{+5}_{-6}$ |
| | | 0.14262 | 0.14144 | 0.58 $^{+12}_{-12}$ | 0.34 $^{+19}_{-19}$ | 0.46 $^{+7}_{-8}$ |
| | | | 0.14226 | 0.58 $^{+13}_{-15}$ | 0.30 $^{+23}_{-23}$ | 0.50 $^{+8}_{-9}$ |
| | | $\kappa_{crit}$ | $\kappa_s$ | 0.58 $^{+8}_{-9}$ | 0.36 $^{+15}_{-15}$ | 0.45 $^{+5}_{-6}$ |
| (1,0,0) | (0,1,0) | 0.14144 | 0.14144 | 0.61 $^{+2}_{-2}$ | 0.40 $^{+6}_{-6}$ | 0.63 $^{+3}_{-3}$ |
| | | | 0.14226 | 0.55 $^{+3}_{-3}$ | 0.34 $^{+8}_{-7}$ | 0.61 $^{+3}_{-3}$ |
| | | 0.14226 | 0.14144 | 0.59 $^{+4}_{-4}$ | 0.34 $^{+9}_{-10}$ | 0.63 $^{+4}_{-4}$ |
| | | | 0.14226 | 0.53 $^{+5}_{-5}$ | 0.28 $^{+12}_{-12}$ | 0.61 $^{+5}_{-6}$ |
| | | 0.14262 | 0.14144 | 0.54 $^{+6}_{-7}$ | 0.23 $^{+15}_{-17}$ | 0.64 $^{+7}_{-6}$ |
| | | | 0.14226 | 0.49 $^{+8}_{-8}$ | 0.20 $^{+20}_{-20}$ | 0.61 $^{+8}_{-8}$ |
| | | $\kappa_{crit}$ | $\kappa_s$ | 0.60 $^{+5}_{-5}$ | 0.36 $^{+12}_{-12}$ | 0.59 $^{+6}_{-6}$ |

Table 9: (cont.)

$$\frac{f_K^+(0)}{Z_V^{eff}} = 0.76 \, {}^{+\,5}_{-\,5} \quad ; \quad m_{1^-}^{c\bar{s}} = 0.67 \, {}^{+\,5}_{-\,3}[a^{-1}]$$

$$\frac{f_K^0(0)}{Z_V^{eff}} = 0.74 \, {}^{+\,5}_{-\,4} \quad ; \quad m_{0^+}^{c\bar{s}} = 0.91 \, {}^{+\,9}_{-\,7}[a^{-1}] \tag{43}$$

The pole mass $m_{1^-}^{c\bar{s}}$, agrees reasonably well with the value $0.732 \, {}^{+\,4}_{-\,4}[a^{-1}]$ quoted in Table 4 and the experimental value $(2.00 \pm 0.12 \pm 0.18)$ GeV [17] quoted in section 2. The result found for $m_{0^+}^{c\bar{s}}$ is also consistent with the value of $2.3 \pm 0.2$ GeV obtained[10] in the lattice simulation of ref. [13]. We also see $f_K^+(0)$ is equal to $f_K^0(0)$ within the errors. These values agree with, and update the preliminary results presented in [52]. In Figs. 2 and 3 we show the form factors $f^+$ and $f^0$ as a function of $q^2$ for two combinations of the light quark masses: $\kappa_l = \kappa_{l_s} = 0.14144$ and $\kappa_l = \kappa_{crit}$, $\kappa_{l_s} = \kappa_s$. In all cases the solid line corresponds to the comparison of the pairs $(q^2, f_K^{+,0}(q^2))$ with the $q^2$-dependence of the form factors determined from a two-parameter pole dominance fit to our data (giving the parameters in eq. (43), for the case $\kappa_l = \kappa_{crit}$, $\kappa_{l_s} = \kappa_s$). Crosses correspond to the form factors at $q^2 = 0$ (up to a factor $Z_V^{eff}$) determined in this way. In the case of $f^+$, we also compare (dashed lines) the $q^2$-dependence of our data with that determined from a one-parameter pole dominance fit, fixing the pole masses to the corresponding values of the vector-meson masses, $m_{1^-}$, quoted in Table 4. Diamonds correspond to the form factors at $q^2 = 0$ (up to a factor $Z_V^{eff}$) determined by using this second method. The fits using the constrained pole masses, also lead to acceptable $\chi^2/dof$. As can be seen from these figures, both methods of extracting $f^+(0)/Z_V^{eff}$ agree remarkably well, which gives us confidence in our procedure[11]. It can be also seen from these figures, that the $q^2$ dependence of both form factors $f_K^+(q^2)$ and $f_K^0(q^2)$ is well described by the pole dominance model.

For the $D \to \pi$ decay[12] we obtain, following the same steps as in the $D \to K$ case:

$$\frac{f_\pi^+(0)}{Z_V^{eff}} = 0.69 \, {}^{+\,10}_{-\,9} \quad ; \quad m_{1^-}^{c\bar{d}} = 0.74 \, {}^{+\,6}_{-\,3}[a^{-1}]$$

---

[10] This value was obtained from the analysis of scalar-scalar two-point functions, not from the study of the $q^2$-dependence of $f_K^0(q^2)$.

[11] Because of our poor determination of the scalar $0^+$ meson mass we can not do a similar comparison for $f^0(0)$. However, the value quoted in eq. (43) for $m_{0^+}^{c\bar{s}}$ agrees well with the value of $2.3 \pm 0.2$ GeV mentioned above ( [13]) and we expect a similar situation for $f^0$ as that obtained for $f^+$.

[12] In order to compute $q^2$ and the form factors in the chiral limit we take $m_\pi = 0.05a^{-1}$ instead of 0, which corresponds to a hopping parameter of $0.14310(2)$ determined from eq. (24). The use of massive instead of massless $u$ and $d$ quarks, which is important for the study of the $q^2$ dependence of the $D \to \pi$ decay, has no practical consequences in the determination of the strange quark mass or in the determination of the $\rho$ meson mass (which has been used in fixing the lattice spacing).

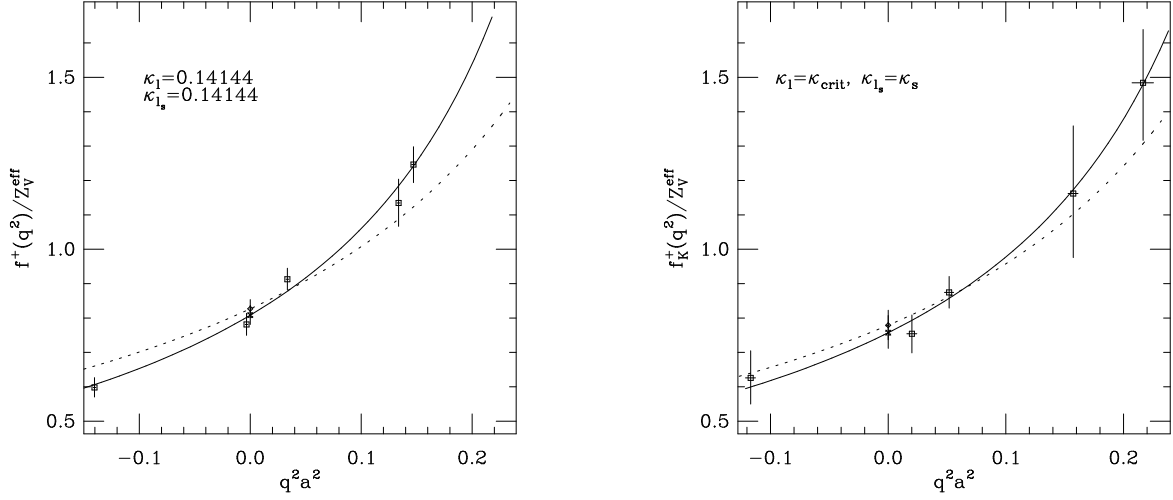

Figure 2: Results for the form factor $f_K^+(q^2)$ as a function of the dimensionless quantity $q^2 a^2$. Left: $\kappa_l = \kappa_{l_s} = 0.14144$. Right: $\kappa_l = \kappa_{crit}, \kappa_{l_s} = \kappa_s$. Solid lines represent the pole dominance behaviour determined from a two-parameter fit to the data (parameters of eq. (43), for the case $\kappa_l = \kappa_{crit}, \kappa_{l_s} = \kappa_s$). Dashed lines represent the pole dominance behaviour determined from a one-parameter fit to the data (fixing the pole masses to the corresponding values of the vector-meson masses, $m_{1^-}$, quoted in Table 4). Crosses and diamonds correspond to the values of the form factor at $q^2 = 0$ (up to a factor $Z_V^{eff}$) determined from two-parameter and one-parameter fits to the data, respectively.

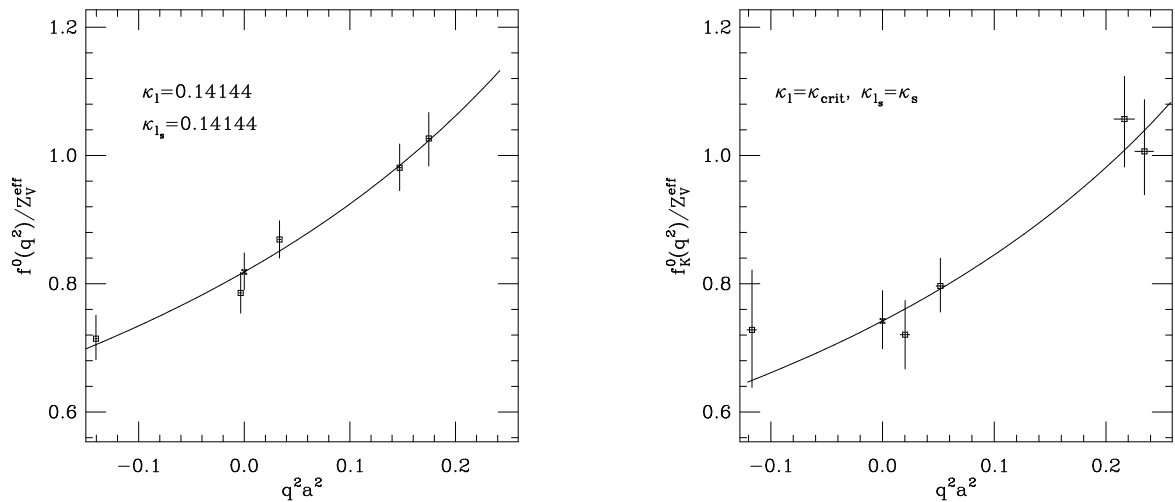

Figure 3: Results for the form factor $f_K^0(q^2)$ as a function of the dimensionless quantity $q^2 a^2$. Left: $\kappa_l = \kappa_{l_s} = 0.14144$. Right: $\kappa_l = \kappa_{crit}, \kappa_{l_s} = \kappa_s$. The curves represent the pole dominance behaviour determined from a two-parameter fit to the data (parameters of eq. (43), for the case $\kappa_l = \kappa_{crit}, \kappa_{l_s} = \kappa_s$). Crosses correspond to the values of the form factor at $q^2 = 0$ (up to a factor $Z_V^{eff}$) determined from a two-parameter fit to the data.

$$\frac{f_\pi^0(0)}{Z_V^{eff}} = 0.60 \ ^{+10}_{-\ 9} \quad ; \quad m_{0+}^{c\bar{d}} = 0.91 \ ^{+13}_{-\ 8} \ [a^{-1}] \qquad (44)$$

$$\frac{f_\pi^+(0)}{f_K^+(0)} = 0.92 \pm 0.08 \qquad (45)$$

where we can see that the value we have obtained for the ratio of form factors $f_\pi^+(0) \ /f_K^+(0)$ is consistent with the experimental numbers quoted in section 2. Note that with the present errors, neither the theoretical nor the experimental result for this ratio gives clear evidence for $SU(3)$-flavour violations (deviations from unity). Note also that $f_\pi^+(0)$ agrees within errors with $f_\pi^0(0)$. In Figure 4 we compare the chirally extrapolated pairs $(q^2, f_\pi^+(q^2))$ with the pole dominance behaviour determined by the parameters of eq. (44) (solid line) and with the results from a one-parameter pole dominance fit, fixing the pole mass to that of the vector-meson mass $m_{1-}^{c\bar{d}}$, quoted in Table 4 (dashed line). Both procedures of extracting $f_\pi^+(0) \ /Z_V^{eff}$ lead again to values for $f_\pi^+(0)$ in an excellent agreement (cross and diamond in Figure 4).

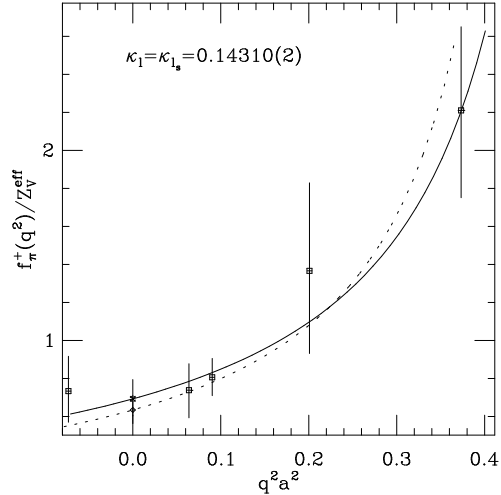

Figure 4: Results for the form factor $f_\pi^+(q^2)$ as a function of the dimensionless quantity $q^2 a^2$. The solid line represents the pole dominance behaviour determined by the parameters of eq. (44). The dashed line corresponds to a one-parameter fit to the data (fixing the pole mass to that of the vector-meson $m_{1-}^{c\bar{d}}$, quoted in Table 4). The cross and the diamond correspond to the values of the form factor at $q^2 = 0$ (up to a factor $Z_V^{eff}$) determined from two-parameter and one-parameter fits to the data, respectively.

## 6.2 The $0^- \to 1^-$ case:

For the $D \to K^*$ decay we obtain:

$$\frac{A_1(0)}{Z_A^{eff}} = 0.67 \, {}^{+\,4}_{-\,4} \quad ; \quad m_{1^+}^{c\bar{s}} = 1.1 \, {}^{+\,3}_{-\,2} \, [a^{-1}] \tag{46}$$

$$\frac{A_2(0)}{Z_A^{eff}} = 0.62 \, {}^{+\,9}_{-\,10} \quad ; \quad m_{1^+}^{c\bar{s}} = 0.46 \, {}^{+\,16}_{-\,7} \, [a^{-1}] \tag{47}$$

$$\frac{A_0(0)}{Z_A^{eff}} = 0.71 \, {}^{+\,5}_{-\,5} \quad ; \quad m_{0^-}^{c\bar{s}} = 0.59 \, {}^{+\,6}_{-\,5} \, [a^{-1}] \tag{48}$$

$$\frac{V(0)}{Z_V^{eff}} = 1.15 \, {}^{+\,7}_{-\,8} \quad ; \quad m_{1^-}^{c\bar{s}} = 0.85 \, {}^{+\,24}_{-\,15} \, [a^{-1}] \tag{49}$$

The results obtained for $m_{1^-}^{c\bar{s}}$ and $m_{0^-}^{c\bar{s}}$ from the form factors $V$ and $A_0$ are consistent within 1 or 1.5 standard deviations with the values quoted in Table 4. In the $A_1$ case, the pole mass is compatible with the value of 2.5 GeV, corresponding to the $D_{s1}$ resonance[13], which was used in the extraction of the form factors in refs. [22]–[23].

In Figures 5–8 we show the form factors $A_1, A_2, A_0$ and $V$ as functions of $q^2$ for two combinations of the light quark masses: $\kappa_l = \kappa_{l_s} = 0.14144$ and $\kappa_l = \kappa_{crit}$, $\kappa_{l_s} = \kappa_s$. In all cases the solid line corresponds to the comparison of the pairs ($q^2$, form factor) with the $q^2$-dependence of the form factors determined from a two-parameter pole dominance fit to our data (parameters of eqs. (46)–(49), for the case $\kappa_l = \kappa_{crit}$, $\kappa_{l_s} = \kappa_s$). Crosses correspond to the form factors at $q^2 = 0$ (up to a factor $Z_V^{eff}$ or $Z_A^{eff}$) determined in this way. In the cases of $V$ and $A_0$, we also compare (dashed lines) the $q^2$-dependence of our data with that determined from a one-parameter pole dominance fit, fixing the pole masses to the corresponding values of the vector and pseudoscalar-meson masses, $m_{1^-}$ and $m_{0^-}$, quoted in Table 4. For the axial form factors $A_1$ and $A_2$, we only make such a comparison for the physical situation $\kappa_l = \kappa_{crit}$, $\kappa_{l_s} = \kappa_s$ where we fix the pole mass to the value used in [22]–[23] (2.5 GeV $\approx 0.9[a^{-1}]$). Diamonds correspond to the form factors at $q^2 = 0$ (up to a factor $Z_V^{eff}$ or $Z_A^{eff}$) determined by using this second method. As can be seen from these figures, both methods of extracting the form factors a $q^2 = 0$ agree well, serving as further check of consistency.

The $q^2$ dependence of $A_0$ and $A_1$ is reasonably well described by the pole dominance model, in contrast with sum-rules calculations which predict for $A_1$ a much weaker $q^2$ dependence than would be given by dominance of the lowest expected $c\bar{s}$ state in the $J^P = 1^+$ channel [37]. However, after our discussion in section 5 of the possible dependence on $q^2$ of the discretisation errors, we must be cautious in our affirmations about the $q^2$ dependence of the form factors and we can not draw any definitive conclusion, without a better understanding of the size and the $q^2$ dependence of the lattice artifacts present in our simulation. In the cases of $A_2$ and $V$, our errors are too large to determine, in a precise way, its $q^2$ dependence.

---

[13]Note however, that the spin-parity quantum numbers of this resonance have not been confirmed yet.

As mentioned above, the values of the form factors at $q^2 = 0$ have been extracted by fitting the chirally-extrapolated data to the pole dominance model. However, in our study we have some momentum channels with values of $q^2$ close to $q^2 = 0$. Another way of obtaining the form factors at $q^2 = 0$ is to take the momentum channel which provides (in the chiral limit) the value of $q^2$ nearest to zero, and by means of the pole dominance model (with a fixed pole mass) extrapolate the form factor to $q^2 = 0$. This method for extracting the form factors at $q^2 = 0$ has the advantage that it only requires a small extrapolation in $q^2$, but on the negative side it only uses a single lattice point. Except in the cases of $V(q^2)$ and $A_1(q^2)$, the results obtained in such a way would agree within errors with those quoted in eqs. (43), (44) and (46–49). For $A_1$ and particularly for $V$, the point nearest to $q^2 = 0$ appears to be high compared to the neighbouring points, giving a higher value of the form factors at $q^2 = 0$ if only this point is used. In the D-decay into vector mesons, the momentum channel with the value of $q^2$ nearest to zero corresponds to the transition $\vec{p}_D = (0, 0, 0) \to |\vec{p}_{K^*}| a = \pi/12$ and is averaged over the six equivalent momenta of the light meson. For $V(q^2)$, the other three channels plotted in Fig. 8, all correspond to transitions in which $\vec{p}_D a = (1, 0, 0)\pi/12$. The difference between these two-sets of points ($\vec{p}_D = \vec{0}$ and $\vec{p}_D a = (1, 0, 0)\pi/12$) is partly statistical but it may also be partly due to systematic errors affecting the two data sets differently. We have decided to be cautious, and to include this difference in the errors in our final results for $A_1(0)/Z_A^{eff}$ and $V(0)/Z_V^{eff}$. Thus, our final values for these two form factors are:

$$\frac{A_1(0)}{Z_A^{eff}} = 0.67 \, ^{+6}_{-4} \tag{50}$$

$$\frac{V(0)}{Z_V^{eff}} = 1.15 \, ^{+28}_{-8} \tag{51}$$

For the $D \to \rho$ decay we find:

$$\frac{A_1^\rho(0)}{Z_A^{eff}} = 0.60 \, ^{+5}_{-4} \; ; \; m_{1^+}^{c\bar{d}} = 1.1 \, ^{+3}_{-2} \, [a^{-1}] \tag{52}$$

$$\frac{A_2^\rho(0)}{Z_A^{eff}} = 0.48 \, ^{+10}_{-11} \; ; \; m_{1^+}^{c\bar{d}} = 0.44 \, ^{+9}_{-5} \, [a^{-1}] \tag{53}$$

$$\frac{A_0^\rho(0)}{Z_A^{eff}} = 0.66 \, ^{+5}_{-5} \; ; \; m_{0^-}^{c\bar{d}} = 0.60 \, ^{+7}_{-5} \, [a^{-1}] \tag{54}$$

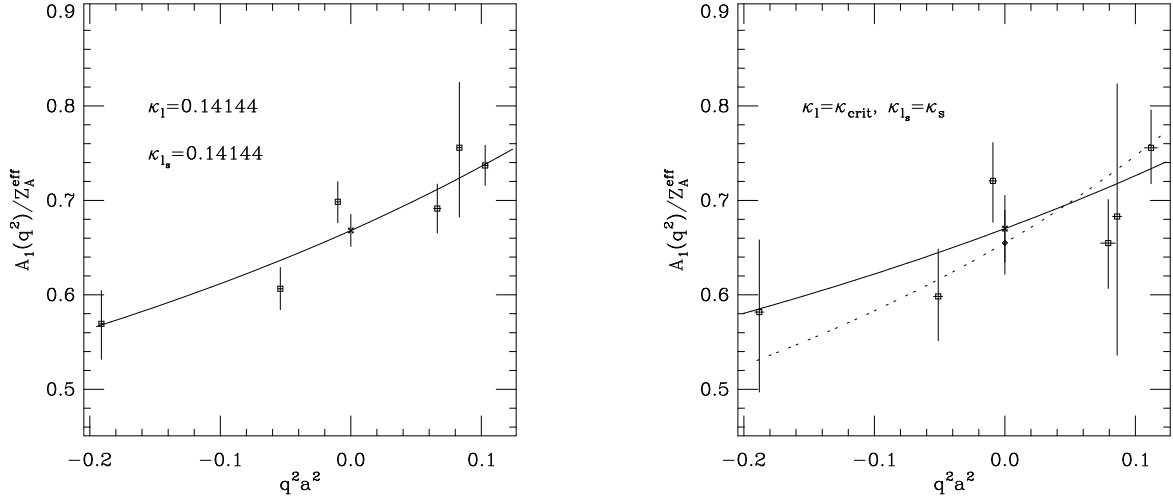

Figure 5: Results for the form factor $A_1(q^2)$ as a function of the dimensionless quantity $q^2 a^2$. Left: $\kappa_l = \kappa_{l_s} = 0.14144$. Right: $\kappa_l = \kappa_{crit}, \kappa_{l_s} = \kappa_s$. Solid lines represent the pole dominance behaviour determined from a two-parameter fit to the data (parameters of eq. (46), for the case $\kappa_l = \kappa_{crit}, \kappa_{l_s} = \kappa_s$). The dashed line, on the right, represents the pole dominance behaviour determined from a one-parameter fit to the data (fixing the pole mass to $0.9[a^{-1}] \approx 2.5\ GeV$). Crosses and diamond correspond to the values of the form factor at $q^2 = 0$ (up to a factor $Z_A^{eff}$) determined from two-parameter and one-parameter fits to the data, respectively.

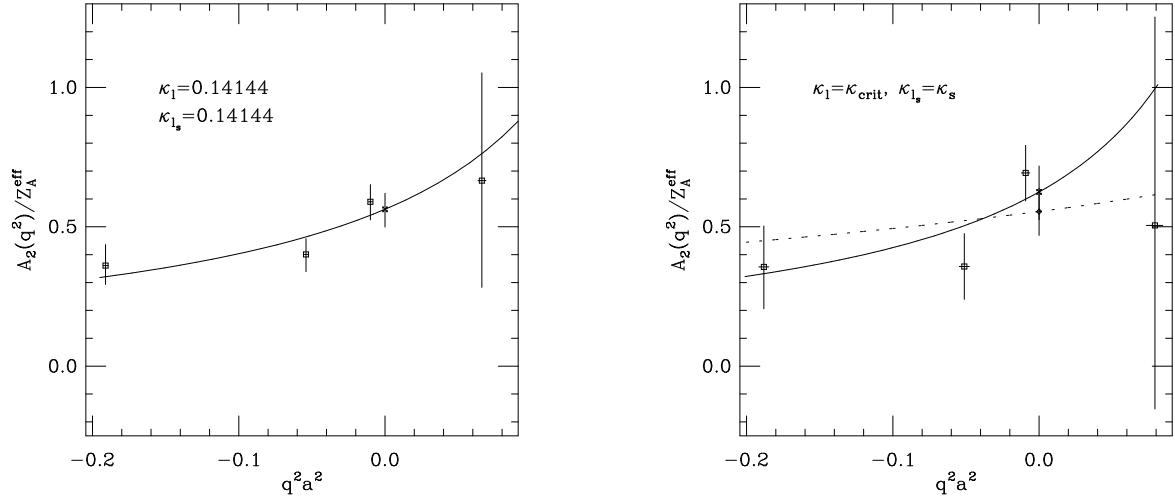

Figure 6: Results for the form factor $A_2(q^2)$ as a function of the dimensionless quantity $q^2 a^2$. Left: $\kappa_l = \kappa_{l_s} = 0.14144$. Right: $\kappa_l = \kappa_{crit}, \kappa_{l_s} = \kappa_s$. Solid lines represent the pole dominance behaviour determined from a two-parameter fit to the data (parameters of eq. (47), for the case $\kappa_l = \kappa_{crit}, \kappa_{l_s} = \kappa_s$). The dashed line, on the right, represents the pole dominance behaviour determined from a one-parameter fit to the data (fixing the pole mass to $0.9[a^{-1}] \approx 2.5\ GeV$). Crosses and diamond correspond to the values of the form factor at $q^2 = 0$ (up to a factor $Z_A^{eff}$) determined from two-parameter and one-parameter fits to the data, respectively.

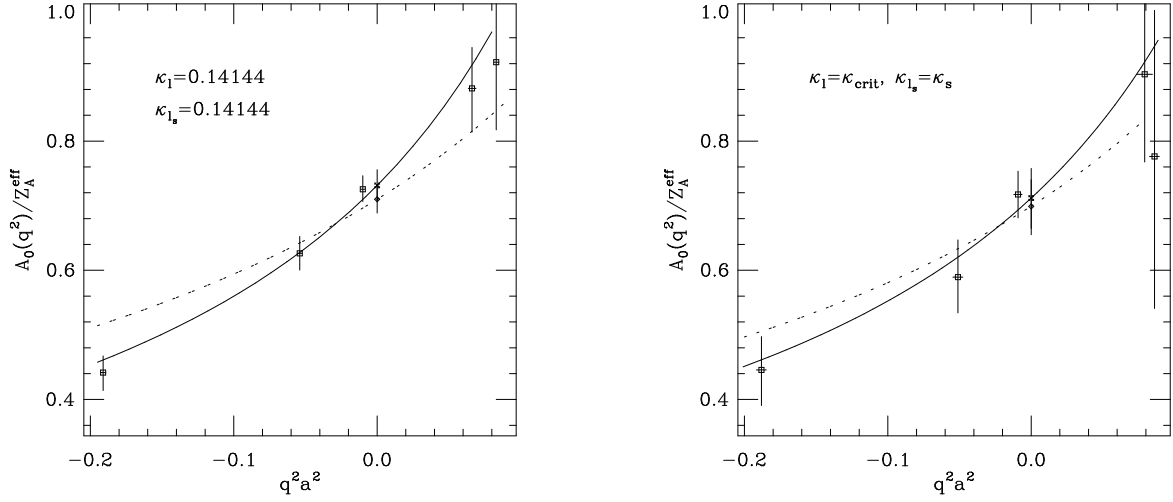

Figure 7: Results for the form factor $A_0(q^2)$ as a function of the dimensionless quantity $q^2 a^2$. Left: $\kappa_l = \kappa_{l_s} = 0.14144$. Right: $\kappa_l = \kappa_{crit}, \kappa_{l_s} = \kappa_s$. Solid lines represent the pole dominance behaviour determined from a two-parameter fit to the data (parameters of eq. (48), for the case $\kappa_l = \kappa_{crit}$, $\kappa_{l_s} = \kappa_s$). Dashed lines represent the pole dominance behaviour determined from a one-parameter fit to the data (fixing the pole masses to the corresponding values of the pseudoscalar-meson masses, $m_{0^-}$, quoted in Table 4). Crosses and diamonds correspond to the values of the form factor at $q^2 = 0$ (up to a factor $Z_A^{eff}$) determined from two-parameter and one-parameter fits to the data, respectively.

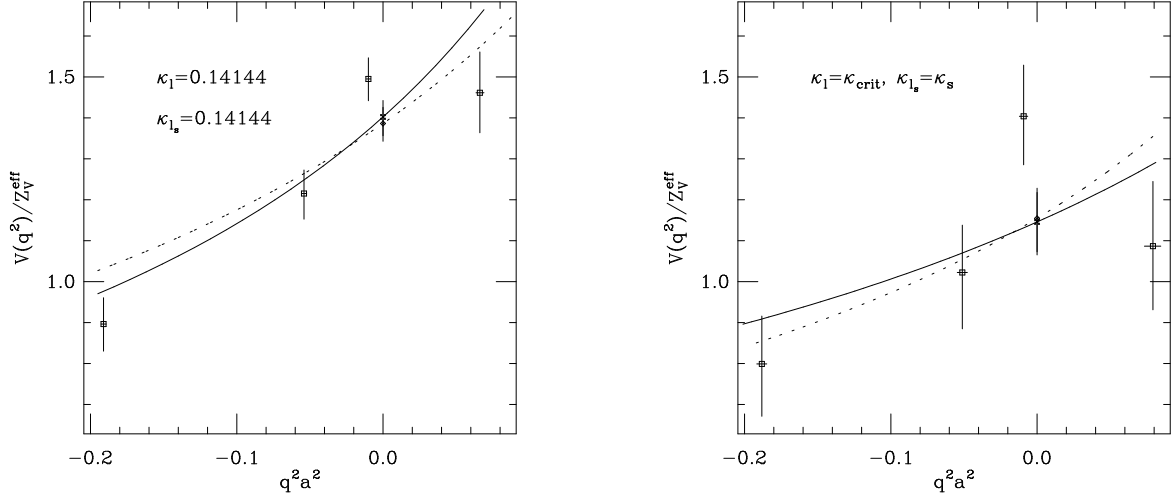

Figure 8: Results for the form factor $V(q^2)$ as a function of the dimensionless quantity $q^2 a^2$. Left: $\kappa_l = \kappa_{l_s} = 0.14144$. Right: $\kappa_l = \kappa_{crit}, \kappa_{l_s} = \kappa_s$. Solid lines represent the pole dominance behaviour determined from a two-parameter fit to the data (parameters of eq. (49), for the case $\kappa_l = \kappa_{crit}, \kappa_{l_s} = \kappa_s$). Dashed lines represent the pole dominance behaviour determined from a one-parameter fit to the data (fixing the pole masses to the corresponding values of the vector-meson masses, $m_{1^-}$, quoted in Table 4). Crosses and diamonds correspond to the values of the form factor at $q^2 = 0$ (up to a factor $Z_V^{eff}$) determined from two-parameter and one-parameter fits to the data, respectively.

$$\frac{V^\rho(0)}{Z_V^{eff}} = 1.08\ ^{+27}_{-10}\ ;\ m_{1^-}^{c\bar{d}} = 0.91\ ^{+36}_{-18}\ [a^{-1}] \tag{55}$$

As in the case of the $D \to K^*$ decay, we have a good determination of the $A_1$ (which dominates the decay rate) and $A_0$ form factors and a poorer determination of $V$ and $A_2$. We have increased the upper errors of the form factors $V$ and $A_1$ by 0.18 and 0.01 respectively, in order to make our quoted values for the form factors at $q^2 = 0$ compatible with the determination of these two form factors, from the momentum channel $\vec{p}_D = (0,0,0) \to |\vec{p}_{K^*}| = \pi/12[a^{-1}]$. In Figure 9 we compare the chirally extrapolated pairs $(q^2, A_1^\rho(q^2))$ with the pole dominance behaviour determined by the parameters of eq. (52).

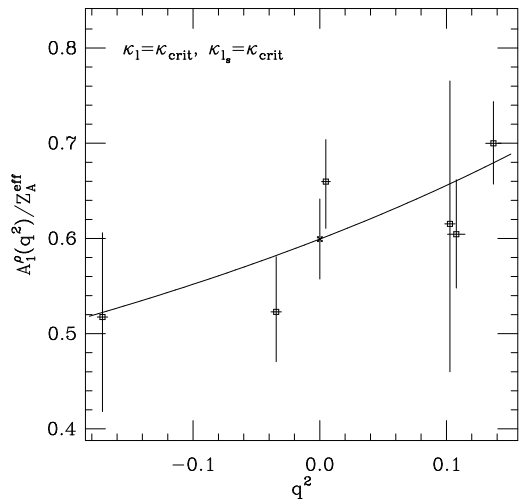

Figure 9: Results for the form factor $(q^2, A_1^\rho(q^2))$ as a function of the dimensionless quantity $q^2 a^2$. The curve represents the pole dominance behaviour determined by the parameters of eq. (52). The cross corresponds to the value of the form factor at $q^2 = 0$ (up to a factor $Z_A^{eff}$) determined from a two-parameter pole dominance fit to the data.

# 7  Conclusions and comparison with experimental data and other calculations

In this section we compare our results with the experimental measurements and other theoretical calculations. In order to do this, we have to specify the values of the effective renormalisation constants $Z_A^{eff}$ and $Z_V^{eff}$; we have taken $Z_A^{eff} = 1.05 \,{}^{+1}_{-8}$ (eq. (42)) and $Z_V^{eff} = 0.88 \,{}^{+4}_{-5}$ (eq. (41)), as discussed in section 5. In Table 10 we show our results for the the semi-leptonic decays $D \to K$ and $D \to K^*$. We have included the above uncertainty in the renormalisation constants in our final results for the form factors in order to account for some of the residual discretisation errors. In our quoted errors for the ratio $A_2/A_1$ we have taken into account the fact that the discretisation errors could be different for different form factors and therefore in general, the effective renormalisation constant $Z_A^{eff}$ could be different for $A_2$ than for $A_1$ giving an additional ambiguity of around 10% for this ratio (see eq. (42)).

In Table 10, we also compare our predictions with the most recent experimental world average and with previous lattice, quark-model and sum-rule results. Our results are in reasonable agreement with the experimental data and the most recent lattice simulations using an $O(a)$-improved SW ([16]) and Wilson ([14]-[15]) actions. Values reported in refs. [11, 12] and [8, 9, 10, 13] (all of them obtained using Wilson-fermions) are also in a good agreement with ours, but the former are in general higher, whereas the latter are smaller, than our predictions. Discretisation errors are, in principle, larger for Wilson than for improved actions, and part of the discrepancies between different lattice results in Table 10, are due to different values used in the literature for the effective renormalisation constants $Z_V^{eff}$ and $Z_A^{eff}$.

Looking now at our result for $f_\pi^+(0)\,/f_K^+(0)$ in eq. (45) there is no clear evidence of $SU(3)$ flavour symmetry breaking and it is consistent with the experimental results. Furthermore, our prediction

$$f_\pi^+(0) = 0.61 \,{}^{+12}_{-11} \tag{56}$$

compares well with lattice calculations obtained with Wilson fermions ($0.58 \pm 0.09$ [10], $0.84 \pm 0.12 \pm 0.35$ [11] and $0.64 \pm 0.09$ [15]) and other theoretical calculations ($0.69$ [30, 33], $0.51$ [32, 34] and $0.6$–$0.75$ [35, 36]). The situation is similar for $f_\pi^0(0)$.

For the decay $D \to \rho$, we find

$$V^\rho(0) = 0.95 \,{}^{+29}_{-14} \quad ; \quad A_1^\rho(0) = 0.63 \,{}^{+6}_{-9} \tag{57}$$

$$A_2^\rho(0) = 0.51 \,{}^{+10}_{-15} \quad ; \quad A_0^\rho(0) = 0.70 \,{}^{+5}_{-12} \tag{58}$$

Our results are in good agreement with the most recent lattice simulation ([15]) and with the previous lattice calculations of refs. [10] and [12] (the results quoted in [10] are however, in general smaller than our predictions) and slightly smaller than the quark model prediction of ref. [30].

In the present study we have found not only the values of the form factors at $q^2 = 0$ for the different decay processes, but also their $q^2$ dependence in a wide region around $q^2 = 0$. Thus we can estimate the integrals of eqs. (8–9) and obtain the total decay rates. Our predictions, together with the experimental measurements and other theoretical calculations, are presented in Table 11. As mentioned above, we have a poor determination of the $q^2$ dependence of the $A_2$ and $V$ form factors in the D decays into vector mesons. However, as can be seen in eq. (9), the contribution of these form factors to the decay rates is small and only important in the proximity of $q^2 = 0$, where their contributions are reasonably well determined. For example, pole masses, for the $A_2$ form factor in the $D \to K^*, \rho$ decays, three times larger than those of eqs. (47,53) give total decay rates and ratios $\Gamma_L/\Gamma_T$ which differ from those quoted in Table 11 only at the level of (0.3-0.5) standard deviations. Therefore, we are confident that we can use the values of the form factors from our simulation for calculating the total decay rates.

| Source | | $f_K^+(0)$ | $f_K^0(0)$ | $V(0)$ | $V/A_1$ |
|---|---|---|---|---|---|
| Exp. | World Ave.[28] | $0.77 \pm 0.04$ | | $1.16 \pm 0.16$ | $1.90 \pm 0.25$ |
| | World Ave.[29] | $0.70 \pm 0.03$ | | | |
| Lattice | This work | $0.67\,^{+7}_{-8}$ | $0.65 \pm 0.07$ | $1.01\,^{+30}_{-13}$ | $1.4\,^{+5}_{-2}$ |
| Gauge | ELC [13] | $0.60\,^{+15}_{-15}\,^{+7}_{-7}$ | | $0.86 \pm 0.24$ | $1.30 \pm 0.2$ |
| | APE [16] | $0.72 \pm 0.09$ | | $1.0 \pm 0.2$ | $1.6 \pm 0.3$ |
| | BKS [11]–[12] | $0.90\,^{+8}_{-8}\,^{+21}_{-21}$ | $0.70\,^{+8}_{-8}\,^{+24}_{-24}$ | $1.43\,^{+45}_{-45}\,^{+48}_{-49}$ | $1.99\,^{+22}_{-22}\,^{+31}_{-35}$ |
| | BG [15] | $0.73 \pm 0.05$ | $0.73 \pm 0.04$ | $1.24 \pm 0.08$ | $1.79 \pm 0.09$ |
| | WU [14] | $0.76 \pm 0.15$ | $0.75 \pm 0.06$ | $1.05 \pm 0.33$ | |
| | LMMS [8]–[10] | $0.63 \pm 0.08$ | | $0.86 \pm 0.10$ | $1.6 \pm 0.2$ |
| Quark | ISGW [32] | $0.76 - 0.82$ | | $1.1$ | $1.4 \pm 0.4$ |
| Models | WSB [30] | $0.76$ | | $1.27$ | $1.4$ |
| | KS [33] | $0.76$ | | $0.8$ | $1.0$ |
| | GS [34] | $0.69$ | | $1.5$ | $2.0$ |
| Sum Rules | BBD [37] | $0.60\,^{+15}_{-10}$ | | $1.10 \pm 0.25$ | $2.2 \pm 0.2$ |
| | AEK [35] | $0.60 \pm 0.15$ | | | |
| | DP [36] | $0.75 \pm 0.05$ | | | |

| Source | | $A_1(0)$ | $A_2(0)$ | $A_2/A_1$ | $A_0$ |
|---|---|---|---|---|---|
| Exp. | World Ave. [28] | $0.61 \pm 0.05$ | $0.45 \pm 0.09$ | $0.74 \pm 0.15$ | |
| Lattice | This work | $0.70\,^{+7}_{-10}$ | $0.66\,^{+10}_{-15}$ | $0.9 \pm 0.2$ | $0.75\,^{+5}_{-11}$ |
| Gauge | ELC [13] | $0.64 \pm 0.16$ | $0.40 \pm 0.28 \pm 0.04$ | $0.6 \pm 0.3$ | |
| | APE [16] | $0.64 \pm 0.11$ | $0.46 \pm 0.34$ | $0.7 \pm 0.4$ | |
| | BKS [12] | $0.83\,^{+14}_{-14}\,^{+28}_{-28}$ | $0.59\,^{+14}_{-14}\,^{+24}_{-23}$ | $0.70\,^{+16}_{-16}\,^{+20}_{-15}$ | $0.94\,^{+9}_{-9}\,^{+22}_{-24}$ |
| | BG [15] | $0.66 \pm 0.03$ | $0.42 \pm 0.17$ | $0.71 \pm 0.20$ | |
| | WU [14] | $0.59 \pm 0.08$ | $0.56 \pm 0.40$ | | |
| | LMMS [8]–[10] | $0.53 \pm 0.03$ | $0.19 \pm 0.21$ | $0.4 \pm 0.4$ | |
| Quark | ISGW [32] | $0.8$ | $0.8$ | $1.0 \pm 0.3$ | |
| Models | WSB [30] | $0.88$ | $1.15$ | $1.3$ | |
| | KS [33] | $0.82$ | $0.8$ | $1.0$ | |
| | GS [34] | $0.73$ | $0.55$ | $0.8$ | |
| Sum Rules | BBD [37] | $0.50 \pm 0.15$ | $0.60 \pm 0.15$ | $1.2 \pm 0.2$ | |

Table 10: Form factors at $q^2 = 0$ for the semi-leptonic decays $D \to K$ and $D \to K^*$: comparison of our results with experimental data and with other theoretical calculations. In obtaining our results we have used $Z_A^{eff} = 1.05\,^{+1}_{-8}$ and $Z_V^{eff} = 0.88\,^{+4}_{-5}$. All lattice gauge calculations have been obtained using Wilson fermions except that of ref. [16] and the present work, where an $O(a)$-improved SW-action has been used.

In the $0^- \to 0^-$ case, $(Z_V^{eff})^2$ is an overall factor in the expression for the width and thus we could quote our result for the decay rate in terms of $(Z_V^{eff})^2/0.88^2$. However, in the decay into vector mesons, the form factors $H^\pm$ mix the contribution of both the vector and the axial form factors and thus such a factorization cannot be made. Therefore in both cases (decays into pseudoscalar and vector mesons) we have decided to include in the quoted statistical errors of our results, the uncertainty due to $Z_V^{eff}$ and $Z_A^{eff}$. We have estimated this uncertainty by computing the extreme values which would be obtained for the different decay rates if the errors of eqs. (41–42) were taken into account. On the other hand, the $q^2$ dependence of the form factors is determined by the different pole masses, quoted in eqs. (43–55), whose physical values depend on the precise value taken for the lattice spacing, $a^{-1}$, and thus the results obtained for the decay rates will also depend on the scale $a^{-1}$. The second set of errors in our results of Table 11 is due to the uncertainty in the determination of the lattice spacing; we have taken $a^{-1} = 2.85 \pm 0.15$ GeV. This ambiguity in the scale has, in general, a small effect on the decay rates, and in some cases is negligible.

As can be seen in Table 11, our results are in excellent agreement with the experimental data. This agreement, together with that already shown in Table 10, provides further confidence that lattice QCD is becoming a reliable quantitative tool for non-perturbative QCD phenomenology. Studies of charm physics on the larger lattices which will shortly become available, will provide a fruitful area of investigation, and will enable the control of the systematic errors (except quenching) present in these calculations. This understanding of the discretisation errors will make it possible to obtain accurate estimates of the the QCD-non perturbative corrections to the $B \to \pi$ and $B \to \rho$ decays, from which we expect to extract the Cabibbo-Kobayashi-Maskawa matrix element $|V_{ub}|$.

We end this paper with a brief summary. The study presented in this paper is one of the first calculations of the form factors of weak vector and axial currents relevant for semi-leptonic decays of D-mesons, performed using the improved quark action proposed by Sheikholeslami and Wohlert [7]. Our results for the form factors (Table 10) and decay rates (Table 11) have reasonably small errors and are in good agreement with experimental measurements. The results at non-zero momentum transfer are, in general, in agreement with the pole dominance model.

| Source | | $\Gamma(D \to K)$ | $\Gamma(D \to K^*)$ | $\Gamma(D \to \pi^\pm)$ | $\Gamma(D \to \rho^\pm)$ |
|---|---|---|---|---|---|
| Exp. | World Ave.[28] | $9.0 \pm 0.5$ | $5.1 \pm 0.5$ | $0.60 \pm 0.15$ | |
| | World Ave.[29] | $7.1 \pm 0.6$ | $4.5 \pm 0.5$ | | |
| Lattice | This work | $7.0 \pm 1.6 \pm 0.4$ | $6.0 {}^{+0.8}_{-1.6}$ | $0.52 \pm 0.18 \pm 0.04$ | $0.43 \pm 0.11$ |
| Gauge | ELC [13] | $5.4 \pm 3.0 \pm 1.4$ | $6.4 \pm 2.8$ | $0.5 \pm 0.3 \pm 0.1$ | $0.60 \pm 0.3 \pm 0.1$ |
| | APE [16] | $7.8 \pm 2.2$ | $6.3 \pm 1.7$ | $0.7 \pm 0.2$ | $0.5 \pm 0.2$ |
| | LMMS [10] | $5.8 \pm 1.5$ | $5.0 \pm 0.9$ | $0.5 \pm 0.2$ | $0.40 \pm 0.09$ |
| Quark | ISGW [32] | $8.5$ | $9.1 \pm 0.25$ | | |
| Models | WSB [30] | $8.26$ | | | |
| | KS [33] | $10.2(e) - 9.9(\mu)$ | | | |
| | GS [34] | $7.1$ | | | |
| Sum | BBD [37] | $6.4 \pm 1.4$ | $3.2 \pm 1.3$ | | |
| Rules | AEK [35] | $5.1 \pm 1.7$ | | | |
| | DP [36] | $8.2 \pm 1.1$ | | $0.76 \pm 0.24$ | |

| Source | | $\frac{\Gamma(D \to K^*)}{\Gamma(D \to K)}$ | $\left(\frac{\Gamma_L}{\Gamma_T}\right)_{K^*}$ | $\left(\frac{\Gamma_L}{\Gamma_T}\right)_\rho$ |
|---|---|---|---|---|
| Exp. | World Ave.[29] | $0.55 \pm 0.07$ | $1.2 \pm 0.1$ | |
| Lattice | This work | $0.86 \pm 0.28 \pm 0.03$ | $1.06 \pm 0.16 \pm 0.02$ | $1.05 {}^{+0.29}_{-0.20} \pm 0.04$ |
| Gauge | ELC [13] | $1.1 \pm 0.6 \pm 0.3$ | $1.4 \pm 0.3$ | |
| | APE [16] | $0.8 \pm 0.3$ | $1.3 \pm 0.3$ | |
| | LMMS [10] | $0.86 \pm 0.22$ | $1.51 \pm 0.27$ | $1.86 \pm 0.56$ |
| Quark | ISGW [32] | $1.1$ | $1.1 \pm 0.2$ | |
| Models | WSB [30] | $1.15$ | $0.9$ | |
| | KS [33] | $0.95$ | $1.1$ | |
| | GS [34] | $1.4$ | $1.2$ | |
| Sum Rules | BBD [37] | $0.5 \pm 0.15$ | $0.86 \pm 0.06$ | |

Table 11: Semi-leptonic partial widths for $D \to K, K^*, \pi$ and $\rho$, using $|V_{cs}| = 0.975$ and $|V_{cd}| = 0.222$. We also report the ratio of the longitudinal to transverse polarisation partial widths for $D \to K^*$ and $D \to \rho$. Units in $10^{10} s^{-1}$. Ref. [28] gives $\Gamma(D \to \pi) = 1.2 \pm 0.3$. We have assumed isospin symmetry and we have taken a value for the decay rate, with charged pions in the final state, of $\Gamma(D \to \pi^\pm) = 0.60 \pm 0.15$.

We have tried to minimize systematics by working with an improved action to reduce discretization errors, and on fairly large volume in the hope that finite-size effects would be small. Nevertheless, it is important that our simulation be repeated on lattices of different sizes and spacings in order to quantify more precisely the systematic effects, which could modify the results presented in this work, in particular the $q^2$-dependence found for the different form factors.

**Acknowledgements** This research was supported by the UK Science and Engineering Research Council under grants GR/G 32779 and GR/J 21347, by the European Union under contract CHRX-CT92-0051, by the University of Edinburgh and by Meiko Limited. We are grateful to Edinburgh University Computing Service and, in particular, to Mike Brown for his tireless efforts in maintaining service on the Meiko i860 Computing Surface. CTS (Senior Fellow), DGR (Advanced Fellow) and NMH acknowledge the support of the Particle Physics and Astronomy Research Council. JN acknowledges the European Union for for their support through the award of a Postdoctoral Fellowship, contract No. CHBICT920066.